\documentclass[preprint,prd,aps,showpacs,showkeys,nofootinbib]{revtex4}
\usepackage{graphicx}
\usepackage{dcolumn}
\usepackage{bm}
\usepackage{ulem}
\usepackage{color}
\usepackage[dvipsnames]{xcolor}
\usepackage{amssymb}
\usepackage{appendix}
\definecolor{light-gray}{gray}{0.78}
\definecolor{mid-gray}{gray}{0.55}
\definecolor{dark-gray}{gray}{0.32}

\begin{document}

\title{The Higgs boson decay $h \rightarrow bs$ in the $U(1)_X$SSM}
\author{Song Gao$^{1,2,3}$, Shu-Min Zhao$^{1,2,3}$\footnote{zhaosm@hbu.edu.cn; 17713155332@163.com}, Ming-Yue Liu$^{1,2,3}$, Xing-Yu Han$^{1,2,3}$, Xi Wang$^{1,2,3}$, Tai-Fu Feng$^{1,2,3,4}$}

\affiliation{$^1$ Department of Physics, Hebei University, Baoding 071002, China}
\affiliation{$^2$ Hebei Key Laboratory of High-precision Computation and Application of Quantum Field Theory, Baoding, 071002, China}
\affiliation{$^3$ Hebei Research Center of the Basic Discipline for Computational Physics, Baoding, 071002, China}
\affiliation{$^4$ Department of Physics, Chongqing University, Chongqing 401331, China}
\date{\today}

\begin{abstract}
In the $U(1)_X$SSM, we delve into the quark flavor violation of $h \rightarrow bs$, where $h$ is identified as the SM-like Higgs boson discovered at the LHC. As the U(1) extension of the minimal supersymmetric standard model (MSSM), the $U(1)_X$SSM has new super fields such as right-handed neutrinos and three Higgs singlets. We conduct a thorough analysis of the underlying mechanisms and parameter dependencies of $h \rightarrow bs$ in the $U(1)_X$SSM. In the $U(1)_X$SSM, we discover that the Br$(h \rightarrow bs)$ for the Higgs decay to $bs$ could significantly differ from the expectation in the standard model (SM), depending on the values of the new parameters introduced in the model. Our research not only contributes to a deeper understanding of Higgs physics within the $U(1)_X$SSM, but also provides valuable guidance for new physics (NP).
\end{abstract}

\maketitle

\section{Introduction}
After the discovery of the Higgs boson\cite{Higg1,NewH1,NewH2,NewH3,NewH4,NewH5,NewH6}, one of the main goals has been to measure its properties with the highest possible precision. Current measurements, within experimental and theoretical uncertainties, are consistent with the predictions of the SM for the Higgs boson. Flavor-changing Higgs interactions provide an intriguing scenario for searching for new physics. In the SM, flavor-changing neutral currents (FCNC) occur only at the loop level and are thus highly suppressed\cite{FCNC}. Therefore, physics beyond the SM may predict larger FCNC compared to the SM. The initial studies of the loop-induced $hbs$ coupling originated in the 1980s\cite{Higg2}, while more recent work based on current data reports that the predicted Br($h \rightarrow bs$) in the SM is approximately $10^{-7}$\cite{Br1,Br2,Br3,Br4}.

 Physicists have expanded the SM, resulting in a number of extended models, among which the Minimal Supersymmetric Standard Model (MSSM) has garnered significant attention. However, There are some problems with MSSM, including the $\mu$-problem\cite{mu1,mu2} and the issue of massless neutrinos\cite{zwz1,zwz2}. To address these problems, we extend the MSSM by introducing the $U(1)_X$ gauge group, with the symmetry group being $SU(3)_C\times SU(2)_L \times U(1)_Y\times U(1)_X$\cite{U1x1,U1x2,U1x3,U1x4}. This extension adds three Higgs singlet superfields and right-handed neutrino superfields to the MSSM\cite{Right1,Right2,Higg3}. In the $U(1)_XSSM$, there is a new gauge boson$(A^{'X}_\mu)$ for the new gauge group $U(1)_X$.
Its neutralino supersymmetric partner is $\lambda_{\tilde{X}}$. Consequently, there are five neutral CP-even Higgs component fields ($\hat{\eta},~\hat{\bar{\eta}},~\hat{S}$, $H_{u},~H_{d}$) in the model,
 and mix together, forming a $5\times 5$ mass-squared matrix. Consequently, the mass of the lightest CP-even Higgs particle can be improved at the tree level.
 In the $U(1)_X$SSM, the small hierarchy problem in the MSSM is alleviated through the added right-handed neutrinos, sneutrinos, and extra Higgs singlets. The $\mu$-problem existing in the MSSM is
  relieved after the spontaneous symmetry breaking of the $S$ field in vacuum through $\lambda_H\hat{S}\hat{H}_u\hat{H}_d$.
   Through the term $Y_\nu\hat{\nu} \hat{l} \hat{H}_u$, the right-handed neutrinos and
 left-handed neutrinos mix together\cite{TREE}, which makes light neutrinos to obtain tiny
 masses through the seesaw mechanism.

Our work is to study quark flavor violating(QFV) decay. In Ref.\cite{QFV}, they study QFV interactions mediated by the Higgs boson $h$. Using an effective field theory approach, they systematically list all possible tree-level ultraviolet completions, which comprise models with vector-like quarks or extra scalars. Based on current limits from low-energy processes, such as meson mixing and $b\rightarrow s\gamma$, upper limits are provided for the allowed flavor-violating transitions. They find that flavor-violating transitions in scenarios with vector-like quarks are always significantly suppressed\cite{LFV11,LFV22}, while a general Two-Higgs-Doublet Model (2HDM) may have a considerable rate\cite{LFV33,LFV44}. They perform comprehensive numerical simulations, taking into account all relevant theoretical and phenomenological constraints. The results show that BR($h\rightarrow bs$) are still allowed at the subpercent (percent) level, which are being explored at the LHC. Meanwhile, the study of lepton flavor violation (LFV) is also crucial. In Refs.\cite{LFV1,LFV2,LFV3}, they study Higgs-mediated flavor-changing neutral current interactions within the effective field theory approach, both with and without adopting the minimal flavor violation(MFV) hypothesis. These works have significantly contributed to our subsequent research on the $h\rightarrow bs$ decay in the $U(1)_X$SSM work.

For flavor-changing couplings of $h\rightarrow bs,~bd$, the LHC experiments have little direct sensitivity\cite{BB1,BB2}. The main decay channel of the Higgs is $h\rightarrow b\bar{b}$, which can mimic $h\rightarrow bs$ by mistagging one of the b-jets. Furthermore, it will not be possible to differentiate between jets from $s$ and $d$ quarks. Therefore, the Higgs decays to $bs$ and $bd$ are effectively indistinguishable. On the other hand, a huge QCD background causes trouble with processes involving bottom quarks. It makes the measurement of decays $h\rightarrow bj,~ (j=s,~d)$ very challenging at the LHC, even at HL-LHC.

To precisely measure the Higgs bosons' properties is one of the ILC's major physics goals. For the branching ratios of Higgs decays $h\rightarrow bs,~bd$, the ILC could theoretically reach subpercent sensitivity\cite{BB3}. Though these channels suffer from large QCD backgrounds and are hard or impossible to see at the LHC, they may be easily accessible in the ILC with clean collider
environment.

In the SM, BR$(h \rightarrow bs)$ is of the order of $10^{-7}$. Through numerical analysis, in the Next-to-Minimal Supersymmetric Standard Model(NMSSM)\cite{NMSSM1,NMSSM2}, BR$(h \rightarrow bs)$ is of the order of $10^{-5}\sim10^{-4}$.
While, BR$(h \rightarrow bs)$in the MSSM is of the order of $10^{-4}$. Comparing with the SM, the new physics particles in
the both models (NMSSM and MSSM) provide a lot
of contribution for the process $h \rightarrow bs$.
In the $U(1)_X$SSM, BR$(h \rightarrow bs)$ can reach the order of $10^{-3}$,
which is obviously larger than the predictions of NMSSM and MSSM.
It implies that the pure UMSSM terms such as $g_X$ and $g_{YX}$ can give large contributions.

The outline of this paper is as follows, in Sec.II, we will introduce the fundamental elements of the $U(1)_X$SSM, including its superpotential, general soft-breaking terms. In Sec.III, we will provide analytical expressions for the branching ratio of the $h \rightarrow bs$ decay in the $U(1)_X$SSM. In Sec.IV, we present the corresponding parameters and numerical analysis. In Sec.V, we offer a summary of this article. Some formulas are shown in the appendix.
\section{The relevant content of $U(1)_X$SSM}

The superpotential of $U(1)_X$SSM is expressed as follows
\begin{eqnarray}
&&W=l_W\hat{S}+\mu\hat{H}_u\hat{H}_d+M_S\hat{S}\hat{S}-Y_d\hat{d}\hat{q}\hat{H}_d-Y_e\hat{e}\hat{l}\hat{H}_d+\lambda_H\hat{S}\hat{H}_u\hat{H}_d
\nonumber\\&&+\lambda_C\hat{S}\hat{\eta}\hat{\bar{\eta}}+\frac{\kappa}{3}\hat{S}\hat{S}\hat{S}+Y_u\hat{u}\hat{q}\hat{H}_u+Y_X\hat{\nu}\hat{\bar{\eta}}\hat{\nu}
+Y_\nu\hat{\nu}\hat{l}\hat{H}_u.
\end{eqnarray}
We use the notations of $v_{u}$, $v_{d}$, $v_{\eta}$, $v_{\bar{\eta}}$ and $v_{S}$ to signify the vacuum expectation values (VEVs) associated with the Higgs superfields $H_{u}$, $H_{d}$, $v_{\eta}$, $v_{\bar{\eta}}$, and $S$, respectively. Subsequently, we present the precise formulations of the two Higgs doublets and three Higgs singlets here
\begin{eqnarray}
&&\hspace{1cm}H_{u}=\left(\begin{array}{c}H_{u}^+\\{1\over\sqrt{2}}\Big(v_{u}+H_{u}^0+iP_{u}^0\Big)\end{array}\right),~~
H_{d}=\left(\begin{array}{c}{1\over\sqrt{2}}\Big(v_{d}+H_{d}^0+iP_{d}^0\Big)\\H_{d}^-\end{array}\right),
\nonumber\\&&\eta={1\over\sqrt{2}}\Big(v_{\eta}+\phi_{\eta}^0+iP_{\eta}^0\Big),~~~
\bar{\eta}={1\over\sqrt{2}}\Big(v_{\bar{\eta}}+\phi_{\bar{\eta}}^0+iP_{\bar{\eta}}^0\Big),~~
S={1\over\sqrt{2}}\Big(v_{S}+\phi_{S}^0+iP_{S}^0\Big).
\end{eqnarray}
And two angles are defined as $\tan\beta$ = $v_{u}$/$v_{d}$ and $\tan\beta_{\eta}$ = $v_{\bar{\eta}}$/$v_{\eta}$.

The soft SUSY breaking terms of $U(1)_X$SSM are as follows
\begin{eqnarray}
&&\mathcal{L}_{soft}=\mathcal{L}_{soft}^{MSSM}-B_SS^2-L_SS-\frac{T_\kappa}{3}S^3-T_{\lambda_C}S\eta\bar{\eta}
+\epsilon_{ij}T_{\lambda_H}SH_d^iH_u^j\nonumber\\&&\hspace{1cm}
-T_X^{IJ}\bar{\eta}\tilde{\nu}_R^{*I}\tilde{\nu}_R^{*J}
+\epsilon_{ij}T^{IJ}_{\nu}H_u^i\tilde{\nu}_R^{I*}\tilde{l}_j^J
-m_{\eta}^2|\eta|^2-m_{\bar{\eta}}^2|\bar{\eta}|^2-m_S^2S^2\nonumber\\&&\hspace{1cm}
-(m_{\tilde{\nu}_R}^2)^{IJ}\tilde{\nu}_R^{I*}\tilde{\nu}_R^{J}
-\frac{1}{2}\Big(M_S\lambda^2_{\tilde{X}}+2M_{BB^\prime}\lambda_{\tilde{B}}\lambda_{\tilde{X}}\Big)+h.c.
\end{eqnarray}
$\mathcal{L}_{soft}^{MSSM}$ represents the soft breaking terms of MSSM.

\begin{table}
\caption{ The superfields in $U(1)_X$SSM}
\begin{tabular}{|c|c|c|c|c|c|c|c|c|c|c|c|}
\hline
Superfields & $\hspace{0.1cm}\hat{q}_i\hspace{0.1cm}$ & $\hat{u}^c_i$ & $\hspace{0.2cm}\hat{d}^c_i\hspace{0.2cm}$ & $\hat{l}_i$ & $\hspace{0.2cm}\hat{e}^c_i\hspace{0.2cm}$ & $\hat{\nu}_i$ & $\hspace{0.1cm}\hat{H}_u\hspace{0.1cm}$ & $\hat{H}_d$ & $\hspace{0.2cm}\hat{\eta}\hspace{0.2cm}$ & $\hspace{0.2cm}\hat{\bar{\eta}}\hspace{0.2cm}$ & $\hspace{0.2cm}\hat{S}\hspace{0.2cm}$ \\
\hline
$SU(3)_C$ & 3 & $\bar{3}$ & $\bar{3}$ & 1 & 1 & 1 & 1 & 1 & 1 & 1 & 1  \\
\hline
$SU(2)_L$ & 2 & 1 & 1 & 2 & 1 & 1 & 2 & 2 & 1 & 1 & 1  \\
\hline
$U(1)_Y$ & 1/6 & -2/3 & 1/3 & -1/2 & 1 & 0 & 1/2 & -1/2 & 0 & 0 & 0  \\
\hline
$U(1)_X$ & 0 & -1/2 & 1/2 & 0 & 1/2 & -1/2 & 1/2 & -1/2 & -1 & 1 & 0  \\
\hline
\end{tabular}
\label{JJ1}
\end{table}

In the context of $U(1)_X$SSM, the simultaneous existence of two Abelian groups, $U(1)_Y$ and $U(1)_X$, gives rise to a new effect that does not exist in MSSM: gauge kinetic mixing. This effect can also be induced through Renormalization Group Equations (RGEs), even if it is assumed as zero at $ M_{GUT}$.

$Y^Y$ denotes the $U(1)_Y$ charge, while $Y^X$ represents the $U(1)_X$ charge. One can write the covariant derivatives of the $U(1)_X$SSM in the following form\cite{Du1,Du2,Du3}
\begin{eqnarray}
&&D_\mu=\partial_\mu-i\left(\begin{array}{cc}Y^Y,&Y^X\end{array}\right)
\left(\begin{array}{cc}g_{Y},&g{'}_{{YX}}\\g{'}_{{XY}},&g{'}_{{X}}\end{array}\right)
\left(\begin{array}{c}A{'}_{\mu}^{Y} \\ A{'}_{\mu}^{X}\end{array}\right)\;,
\end{eqnarray}
where $A_{\mu}^{\prime Y}$ and $A_{\mu}^{\prime X}$ represent the gauge fields of $U(1)_Y$ and $U(1)_X$, respectively.

Under the condition that the two Abelian gauge groups remain unbroken, we employ the rotation matrix R to carry out a change of basis\cite{Du3}
\begin{eqnarray}
&&\left(\begin{array}{cc}g_{Y},&g{'}_{{YX}}\\g{'}_{{XY}},&g{'}_{{X}}\end{array}\right)
R^T=\left(\begin{array}{cc}g_{1},&g_{{YX}}\\0,&g_{{X}}\end{array}\right)~,~~~~
%%%%%%%%%%%%%%%%%%%%%%%%%%%%%%%%%%%%%%%%%%%%%%%%%%%%%%%%%%%%%%%%%%%%
R\left(\begin{array}{c}A_{\mu}^{\prime Y} \\ A_{\mu}^{\prime X}\end{array}\right)
=\left(\begin{array}{c}A_{\mu}^{Y} \\ A_{\mu}^{X}\end{array}\right)\;,
\end{eqnarray}
In the end, the covariant derivatives of the $U(1)_X$SSM transform into
\begin{eqnarray}
&&D_\mu=\partial_\mu-i\left(\begin{array}{cc}Y^Y,&Y^X\end{array}\right)
\left(\begin{array}{cc}g_{1},&g_{{YX}}\\0,&g_{{X}}\end{array}\right)
\left(\begin{array}{c}A_{\mu}^{Y} \\ A_{\mu}^{X}\end{array}\right)\;.
\end{eqnarray}
At the tree level, three neutral gauge bosons $A_{\mu}^{\prime X}$, $A_{\mu}^{\prime Y}$, and $V_{\mu}^{\prime 3}$ mix with each other($V_{\mu}^{\prime 3}$ represents the neutral gauge field of $SU(2)_L$), with their mass matrix represented in the basis of ($A_{\mu}^{\prime Y}$, $V_{\mu}^{\prime 3}$, $A_{\mu}^{\prime X}$).
\begin{eqnarray}
\left(\begin{array}{ccc}\frac{1}{8}g_{1}^{2}v^{2},
&-\frac{1}{8}g_{1}g_{2}v^{2},&\frac{1}{8}g_{1}(g_{YX}+g_{X})v^{2}\\
-\frac{1}{8}g_{1}g_{2}v^{2},
&\frac{1}{8}g_{1}g_{2}v^{2},&-\frac{1}{8}g_{2}(g_{X}+g_{YX})v^{2}\\
(g_{YX}+g_{X})v^{2},&(g_{YX}+g_{X})v^{2},&\frac{1}{8}g_{X}^{2}\xi^{2}\end{array}\right),
\end{eqnarray}
with $v^2$ = $v_u^2$ + $v_d^2$ and $\xi^2$ = $v_\eta^2$ + $v_{\bar\eta}^2$.
Some mass matrices and coupling vertices are shown in Appendix A and Refs.\cite{Right2,Higgs coup,TREE}.
\section{Process Analysis}

\begin{figure}[ht]
\setlength{\unitlength}{4.0mm}
\centering
\includegraphics[width=2.5in]{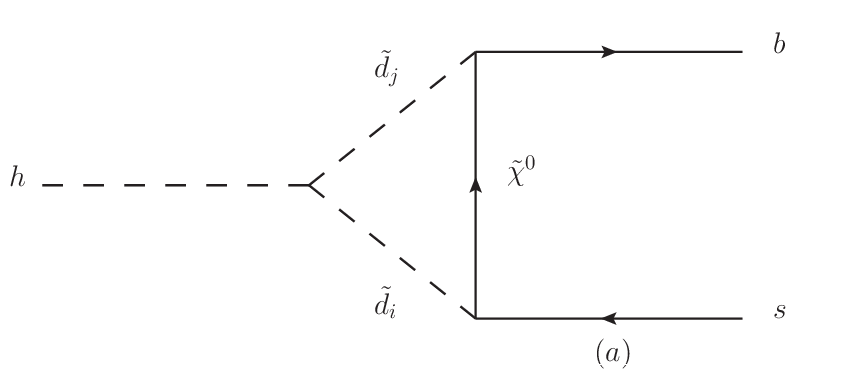}
\includegraphics[width=2.5in]{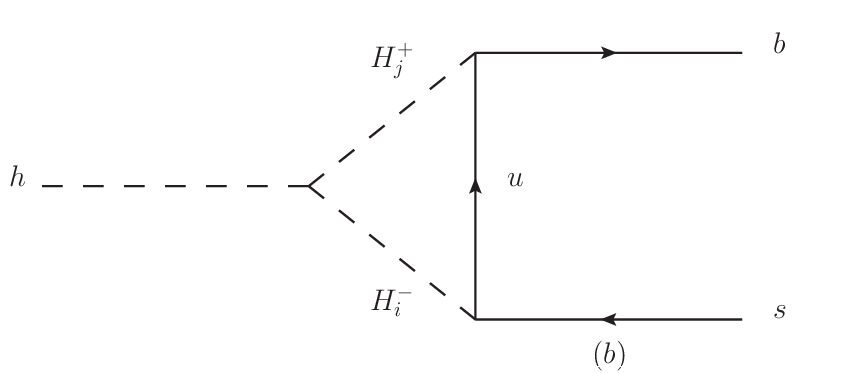}
\includegraphics[width=2.5in]{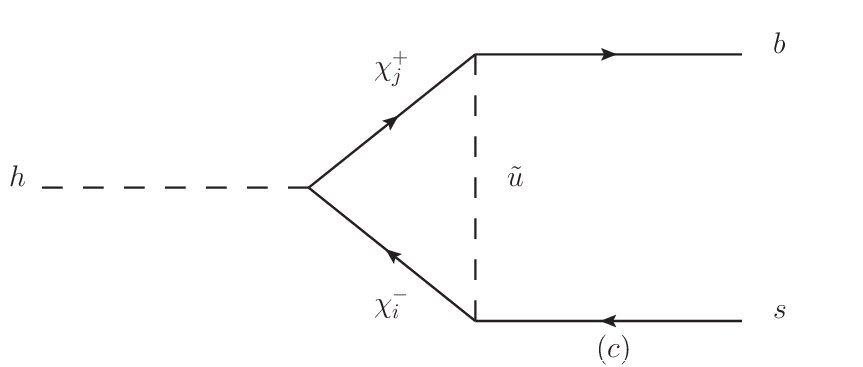}
\includegraphics[width=2.5in]{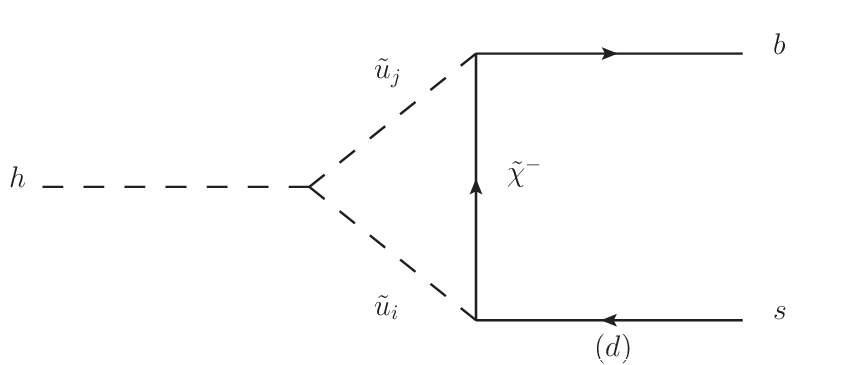}
\includegraphics[width=2.5in]{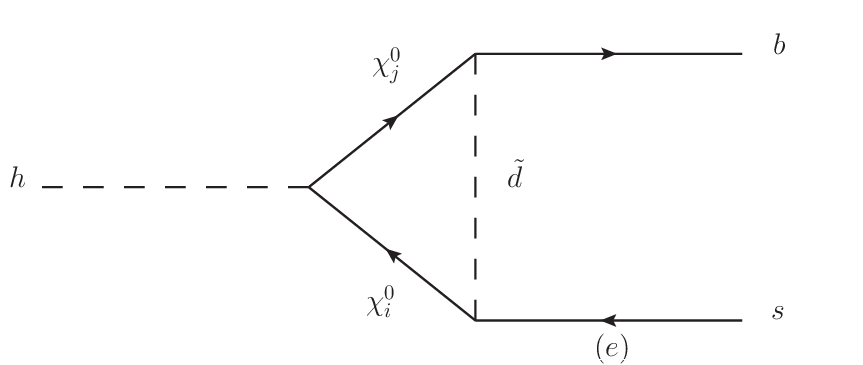}
\includegraphics[width=2.5in]{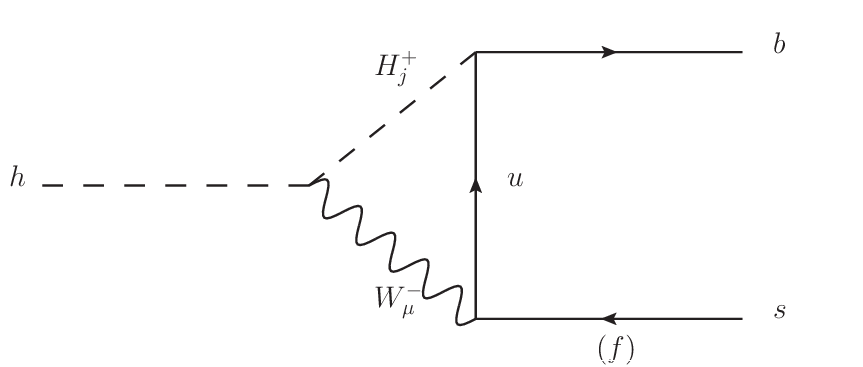}
\includegraphics[width=2.5in]{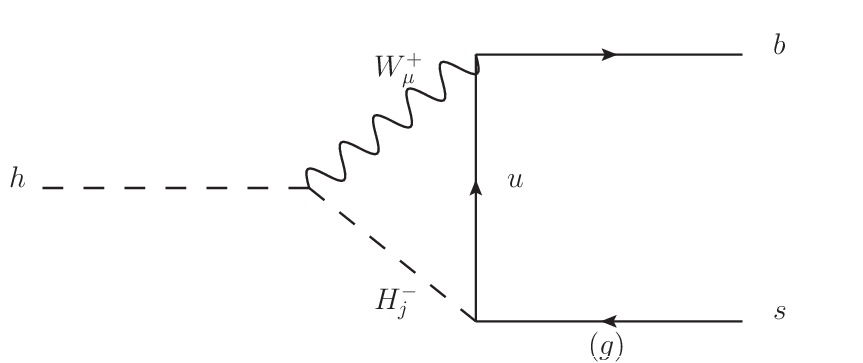}
\includegraphics[width=2.5in]{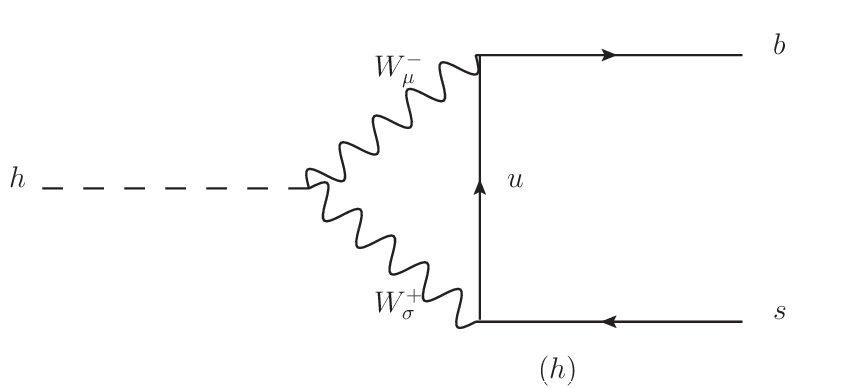}\\
\includegraphics[width=1.5in]{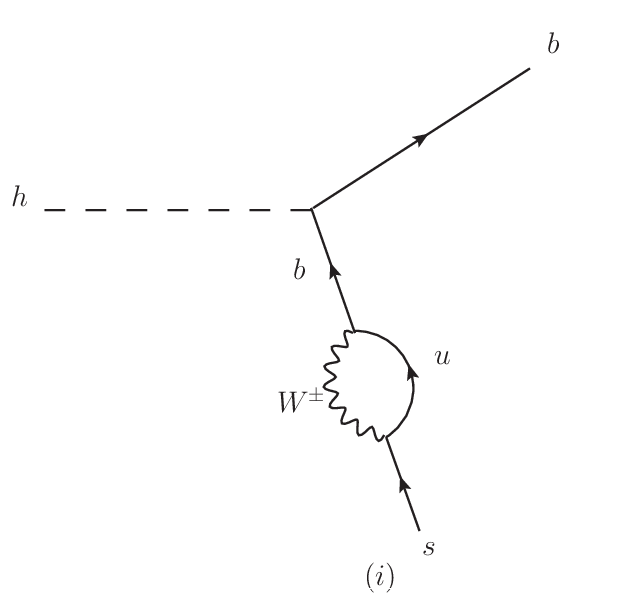}
\includegraphics[width=1.5in]{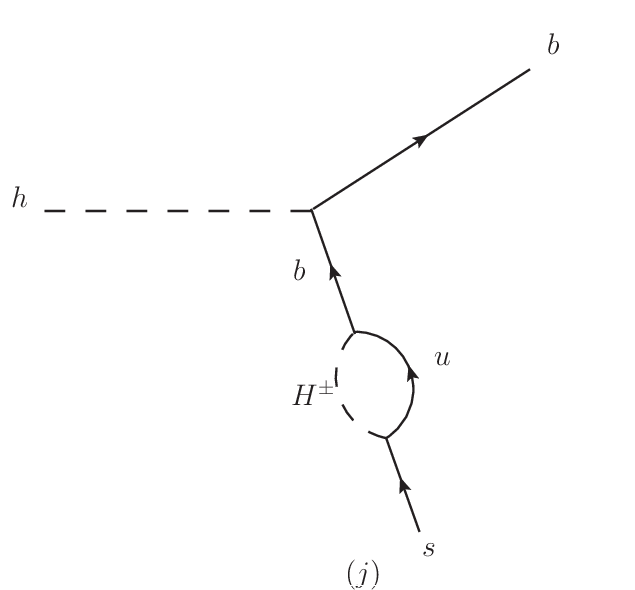}
\includegraphics[width=1.5in]{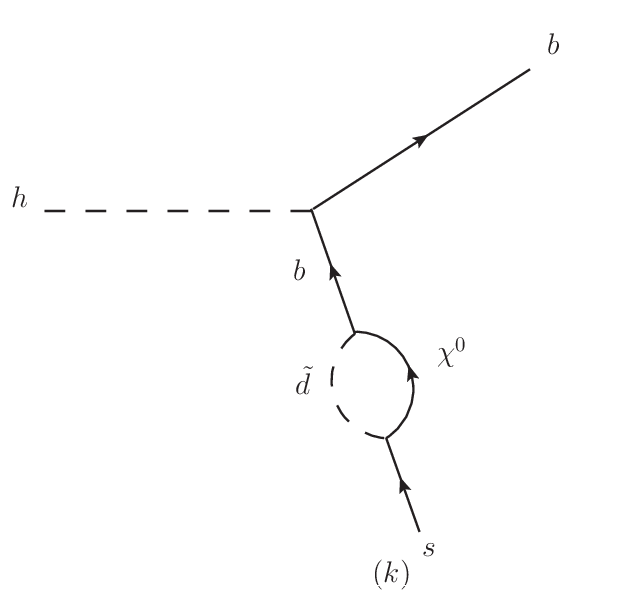}
\includegraphics[width=1.5in]{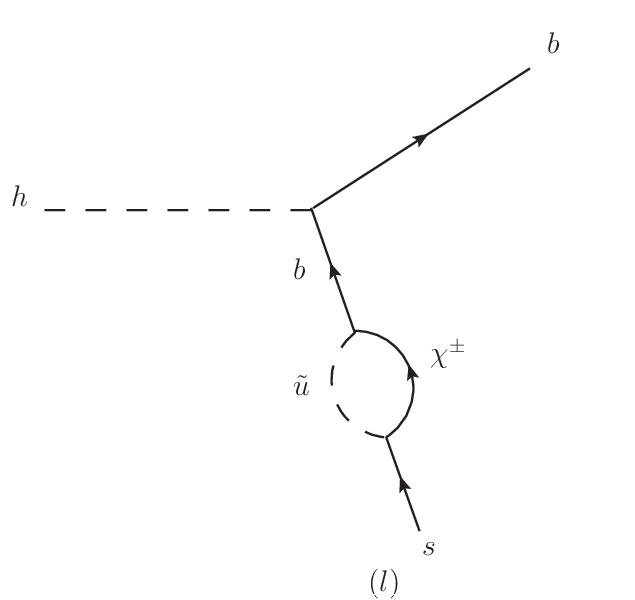}\\
\includegraphics[width=1.5in]{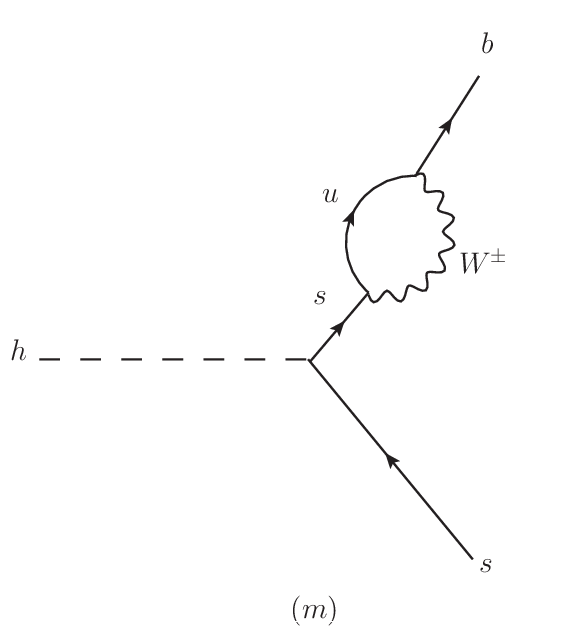}
\includegraphics[width=1.5in]{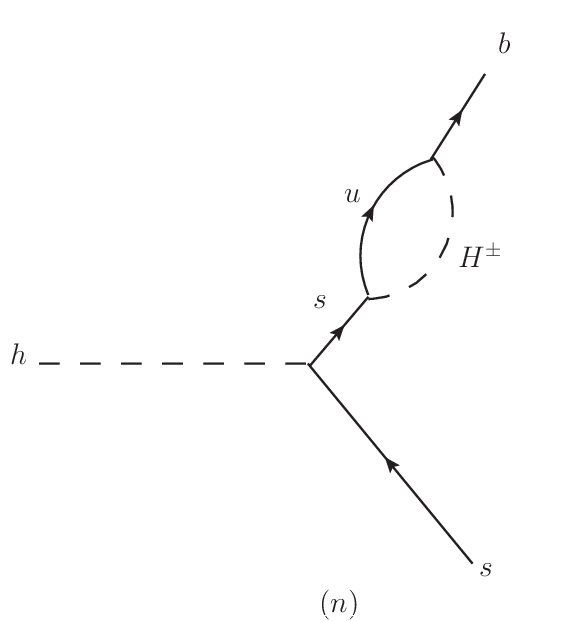}
\includegraphics[width=1.5in]{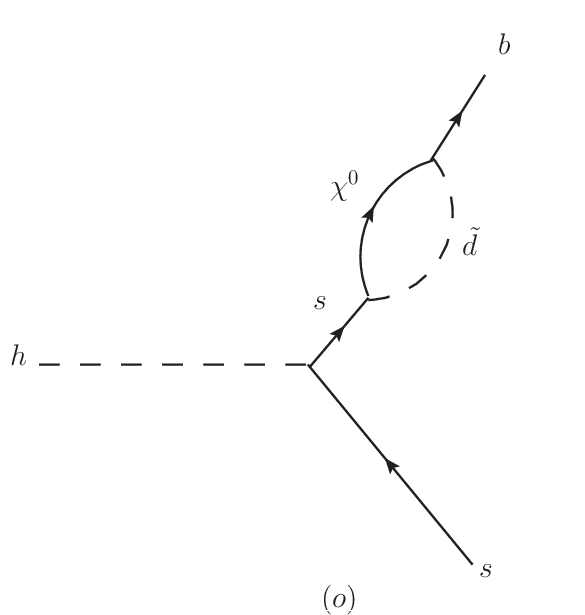}
\includegraphics[width=1.5in]{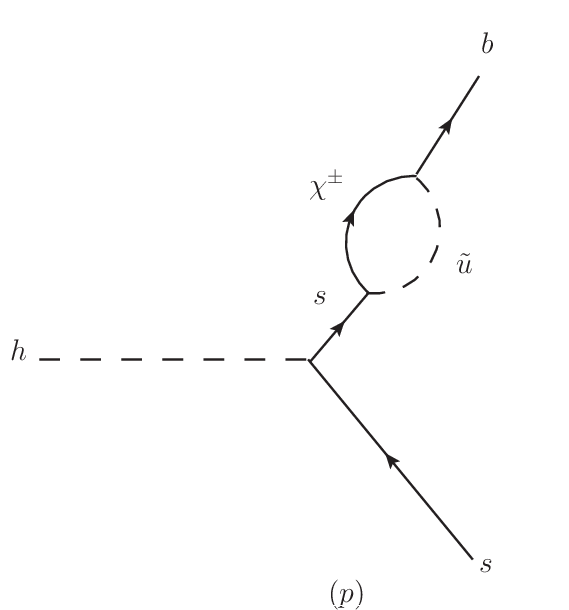}
\caption{  Feynman diagrams for the $h \rightarrow bs$ process in the $U(1)_X$SSM.}\label{N1}
\end{figure}

In this section, we investigate the amplitude and branching ratio of $h \rightarrow bs$. The corresponding Feynman diagrams are depicted in Fig.$\ref{N1}$. For example, we analyze one of the Feynman diagrams in Fig.$\ref{N1}$. The Feynman amplitude of Fig.$\ref{N1}$(a) is

\begin{eqnarray}
&&\mathcal{M}_{(a)}=\bar{U}_b(p)\int\frac{d^Dk}{(2\pi)^D}\frac{1}{[(k+q)^2-m_{{\chi}^0_k}^2][(p+q+k)^2-m_{\tilde{d}_j}^2](k^2-m_{\tilde{d}_i}^2)} \nonumber\\&&\hspace{1.5cm} \times\Big((B_LP_L+B_RP_R)({k\!\!\!\slash+q\!\!\!\slash-m_{{\chi}^0}})(A_LP_L+A_RP_R)\Big)C_1U_s(q).\label{x0}
\end{eqnarray}

In this expression, $p$ represents the momentum of the bottom quark (b), $q$ corresponds to the momentum of the strange quark (s), and $k$ is the loop momentum. The $m_{\tilde{d}_{i(j)}}$ term corresponds to the mass of the down$-$squark. $A_L, B_L, A_R,$ and $B_R$ represent the coupling vertices mentioned in Section II.

 $A_L$ and $A_R$ are left-handed and right-handed couplings of the vertex $\bar{\chi}^0_id_j\tilde{d_k}$:
\begin{eqnarray}
&&A_L = -\frac{1}{6}\Big(\sqrt{2}g_{1}N^*_{i1}Z^D_{ka} - 3\sqrt{2}g_{2}N^*_{i2}Z^D_{ka} + \sqrt{2}g_{YX}N^*_{i5}Z^D_{ka} + 6N^*_{i3}Y_{d,a}Z^D_{k3+a}\Big)\nonumber\\&&
A_R = -\frac{1}{6}\Big(6Y^*_{d,a}Z^D_{kb}N_{i3} + \sqrt{2}Z^D_{k3+a}[2g_{1}N_{i1} + (2g_{YX} + 3g_X)N_{i5}]\Big).
\end{eqnarray}
$B_L$ and $B_R$ are left-handed and right-handed couplings of the vertex $\bar{d}_i\chi^0_j\tilde{d_k}$:
\begin{eqnarray}
&&B_L = -\frac{1}{6}\Big(2\sqrt{2}g_1 N^*_{j1}Z^{D,*}_{k3+a} + \sqrt{2}(2g_{YX} + 3g_X) N^*_{j5}Z^{D,*}_{k3+a} + 6N^*_{j3}Z^{D,*}_{kb}Y_{d,a}\Big)\nonumber\\&&
B_R = -\frac{1}{6}\Big(6Y^*_{d,a}Z^{D,*}_{k3+a}N_{j3} + \sqrt{2}Z^{D,*}_{ka}[-3g_2 N_{j2} + g_1 N_{j1} + g_{YX}N_{j5}]\Big).
\end{eqnarray}
$\bar U_{b}(p)$ and $U_{s}(q)$ represent the wave functions of the bottom and strange quarks. $C_1$ is the coupling constant of $h\tilde{d}_i \tilde{d}_j$. The subscripts L and R represent the left$-$handed and right$-$handed parts, respectively. In order to save space in the text, the detailed calculation process is shown in Appendix B.

The decay width $\Gamma$ of $h \rightarrow bs$ is obtained by substituting $|\mathcal{M}|^2$ into the following formula\cite{FCNC}
\begin{eqnarray}
&&\Gamma(h\rightarrow bs) = 2N_c\frac{\lambda^\frac{1}{2}(m^2_h,m^2_b,m^2_s)}{16\pi m^3_h}|\mathcal{M}|^2,
\end{eqnarray}
here, $N_c$ = 3 is a color factor, $\lambda(x,y,z) = (x - y - z)^2 - 4y^2z^2$.

The branching ratio we obtained is
\begin{eqnarray}
&&Br(h\rightarrow bs) = \frac{\Gamma(h\rightarrow bs)}{\Gamma(h)}.
\end{eqnarray}
Here, $\Gamma(h) = 4.1 \times 10^{-3} \rm{GeV}$\cite{HiggsM}.

\section{Numerical analysis}
In analyzing the numerical values, we consider the experimental constraints to study the flavor-changing neutral current process $h \rightarrow bs$, subject to the following restrictions.

1. The mass of the lightest CP-even Higgs, designated as $m_h$, approximately equals 125.25 GeV\cite{pdg}.
 The parameters used to calculate Br$(h \rightarrow bs)$ satisfy the 3$\sigma$ limit of the Higgs mass. Currently,  the mass of the Higgs boson, with $3\sigma$ level errors, is obtained
\begin{eqnarray}
&&m_h = 125.25 \pm 0.51\mathrm{GeV}.
\end{eqnarray}

2. The Higgs $h$ decays$(h \rightarrow \gamma + \gamma,~ Z + Z,~ W + W,~ b + \bar b,~ \tau + \bar\tau)$[29] should be satisfied.

3. Muon anomalous magnetic dipole moment is also taken into account\cite{g-21,g-22}.

4. According to the latest LHC data\cite{limit1,limit2,limit3,limit4,limit5,limit6}, we take for the scalar lepton mass greater than 600 GeV, the chargino mass greater than 1000 GeV, and the masses of up-squarks and down-squarks greater than 1400 GeV.

In the calcaulations, we take the up quark mass $m_u$ = 0.0022 GeV, the down quark mass $m_d$ = 0.0047 GeV, the strange quark mass $m_s$ = 0.095 GeV, the bottom quark mass $m_b$ = 4.18 GeV, the charm quark mass $m_c$ = 1.275 GeV, the top quark mass $m_t$ = 173.5 GeV, the $W$ boson mass $m_W$ = 80.385 GeV, the Z boson mass $m_Z$ = 91.188 GeV, the electron mass $m_e$ =  0.500 MeV, the muon mass $m_\mu$ = 0.105 GeV, $\alpha(m_Z)$ = 1/128, $\alpha_s(m_Z)$ = 0.118.
\subsection{one-dimensional line graph(in the NMSSM and MSSM)}
\begin{figure}[ht]
\setlength{\unitlength}{5mm}
\centering
\includegraphics[width=2.5in]{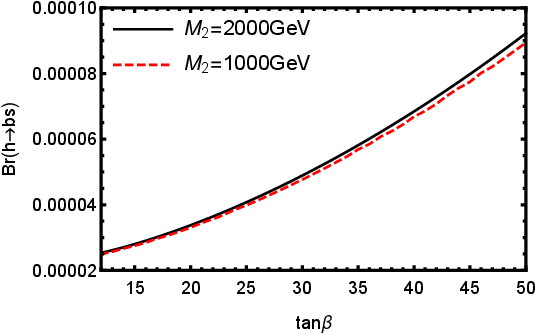}
\caption{The influence of $\tan\beta$ on Br$(h \rightarrow bs)$ in the NMSSM.} {\label {TNMSSM}}
\end{figure}

In the NMSSM, we use images to visualize the effects of variables on the results, where the quantitative parameters are set as follows
\begin{eqnarray}
&&\mu = 0.8 {\rm TeV},~~ v_S = 4.1 {\rm TeV},~~ \kappa = 0.1,~~ \lambda_H = 0.4,~~ M_{\tilde{q}11}^2 = M_{\tilde{q}22}^2 = M_{\tilde{q}33}^2 = 2.5^2 {\rm TeV^2},\nonumber\\
&& M_{\tilde{u}11}^2 = M_{\tilde{d}11}^2 = 2.7^2 {\rm TeV^2},~~M_{\tilde{u}22}^2 = M_{\tilde{d}22}^2 = 2.0^2 {\rm TeV^2},~~ M_{\tilde{u}33}^2 = M_{\tilde{d}33}^2 = 1.5^2 {\rm TeV^2}.
\end{eqnarray}
We conduct numerical calculations for Br$(h \rightarrow bs)$ and plot a correlation graph between $\tan\beta$ and Br$(h \rightarrow bs)$ to clearly illustrate the numerical results.
$\tan\beta$ is a sensitive parameter. In the Fig.$\ref{TNMSSM}$, we plot Br$(h \rightarrow bs)$ versus $\tan\beta$ by the black-solid line($M_2$ = 2000 GeV) and red-dashed line($M_2$ = 1000GeV). We observe that, within the $\tan\beta$ range from 12 to 50, the corresponding values of the lines gradually increase with the increasing $\tan\beta$. The obtained numerical results of Br$(h \rightarrow bs)$ are in the region ($2.5\times10^{-5}\sim9.0\times10^{-5}$).

\begin{figure}[ht]
\setlength{\unitlength}{5mm}
\centering
\includegraphics[width=2.5in]{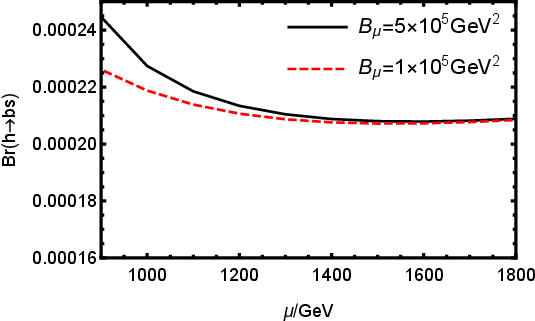}
\caption{The influence of $\mu$ on Br$(h \rightarrow bs)$ in the MSSM.} {\label {TMSSM}}
\end{figure}

In the MSSM, we use images to analyze the impact of variables on the results, with the quantitative parameters as follows
\begin{eqnarray}
&&\mu = 1 {\rm TeV},~~ M_1 = 0.7 {\rm TeV},~~ M_2 = 2 {\rm TeV},~~M_{\tilde{q}11}^2 = M_{\tilde{q}22}^2 = 2.5^2 {\rm TeV^2},\nonumber\\
&&M_{\tilde{d}11}^2 = M_{\tilde{d}22}^2 = M_{\tilde{d}33}^2 = M_{\tilde{u}11}^2= M_{\tilde{u}22}^2 = M_{\tilde{u}33}^2 = M_{\tilde{q}33}^2 = 2.7^2 {\rm TeV^2}.
\end{eqnarray}
We study Br$(h \rightarrow bs)$ numerically and plot diagram between $\mu$ and Br$(h \rightarrow bs)$
to clearly illustrate their relation.
In the Fig.$\ref{TMSSM}$, we plot Br$(h \rightarrow bs)$ versus $\mu$ by the black-solid line($B_{\mu} = 5 \times 10^5$ {GeV}$^2$) and red-dashed line($B_{\mu} = 1 \times 10^5$ {GeV}$^2$). We observe that within the range of $\mu$ from 900 GeV to 1400 GeV, the corresponding value of the line gradually decreases with the increase of $\mu$, and the decreasing speed is getting smaller and smaller. In the range of $\mu$ from 1400 GeV to 1800 GeV, the corresponding value of the line remains almost unchanged. This is because when $\mu$ increases to a certain value, its influence mechanism becomes smaller and smaller. The obtained numerical results of Br$(h \rightarrow bs)$ are in the region ($2.15\times10^{-4}\sim2.45\times10^{-4}$).

\subsection{one-dimensional line graph(in the $U(1)_XSSM$)}
We use images to visualize the effects of variables on the results, where the quantitative parameters are set as follows
\begin{eqnarray}
&&\mu = 0.8 {\rm TeV},~~ M_S = 3.6 {\rm TeV},~~ v_S = 4.3 {\rm TeV},~~ \kappa = 0.1,~~ \lambda_C = -0.08,\nonumber\\
&&M_{\tilde{q}11}^2 = M_{\tilde{q}22}^2 = 1.9^2 {\rm TeV^2},~~ M_{\tilde{u}11}^2 = M_{\tilde{u}22}^2 = 1.9^2 {\rm TeV^2},~~ M_{\tilde{q}33}^2 = M_{\tilde{u}33}^2 = 6 {\rm TeV^2}.
\end{eqnarray}

We use the parameters in Eq.(\ref{a1}), as variables to study the numerical results. To simplify the numerical research, we adopt the assumption $M_{\tilde{d}ij}^2=M_{\tilde{d}ji}^2,~(i,j=1,2,3,~i\neq j).$
\begin{eqnarray}
&&g_X,~~g_{YX},~~tan\beta,~~\lambda_H,~~M_1,~~M_2,\nonumber\\
&&M_{BL},~~M_{BB'},~~M_{\tilde{d}ij}^2=M_{\tilde{d}ji}^2,~(i,j=1,2,3,~i\neq j).\label{a1}
\end{eqnarray}

Typically, the non-diagonal elements of the parameters are set to zero as a default, unless explicitly stated otherwise.

 In this subsection, we conduct numerical calculations for Br$(h \rightarrow bs)$ and plot correlation graphs  for various parameters to clearly illustrate the numerical results. We use the parameters as $\tan\beta$ = 13, $M_1$ = 0.7 {TeV}, $M_2$ = 1.2 {TeV}, $(M_d)^2_{11} = (M_d)^2_{22} = (M_d)^2_{33}$ = 2.70 $\mathrm{TeV}^2$, $M_{BL}$ = 0.7 TeV, $M_{BB'}$ = 0.4 TeV($M_{BL}$ is the mass of the superpartner for the gauge boson under the $U(1)_X$ group.).
  These adopted parameter values are representative, and the obtained branching ratios of $h\rightarrow bs$ are at the order of $10^{-3}$, which is the typical value for the Br expected in this model.
  It may be easier to be measured at HL-LHC and future colliders.

  \begin{figure}[ht]
\setlength{\unitlength}{5mm}
\centering
\includegraphics[width=2.5in]{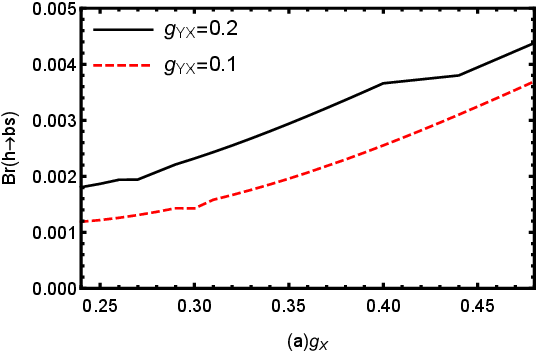}
\setlength{\unitlength}{5mm}
\centering
\includegraphics[width=2.5in]{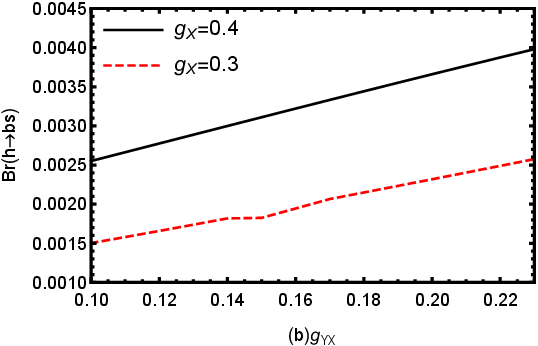}
\setlength{\unitlength}{5mm}
\centering\nonumber\\
\includegraphics[width=2.5in]{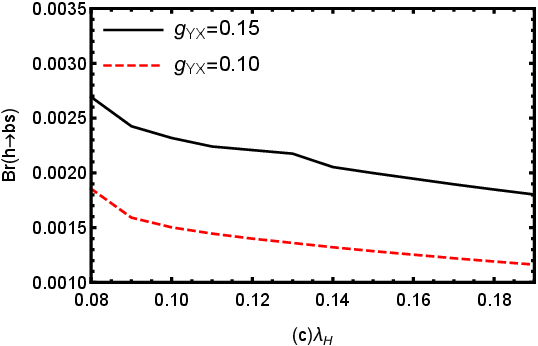}
\caption{The influence of various parameters on Br$(h \rightarrow bs)$: In (a) and (b), $\lambda_H = 0.1$, and in (c), $g_X = 0.3$.} {\label {T1}}
\end{figure}

  $g_X$ is the coupling constant of the new gauge $U(1)_X$.
The mass matrixes of several particles(neutralino, down-squark, up-squark, neutral Higgs, charged Higgs) and several coupling vertices
($h\bar{\chi}_i^0\chi_j^0$ $h\tilde{u}^*_i\tilde{u}_j$ $h\tilde{d}^*_i\tilde{d}_j$
$\bar{\chi}^0_id_j\tilde{d_k}$ $\bar{d}_i\chi^0_j\tilde{d_k}$) all have the important parameter $g_X$, which can improve the new physics effect.
Obviously, $g_X$ should be a sensitive parameter. In the  Fig.$\ref{T1}$ (a),
we plot Br$(h\rightarrow bs)$ versus $g_X$ by the
black-solid line$(g_{YX}=0.2)$ and red-dashed line$(g_{YX}=0.1)$ with $\lambda_H=0.1$. We observe that, within the $g_X$ range from 0.24 to 0.48, the corresponding values of the lines gradually increase with the increasing $g_X$.
The obtained numerical results of Br$(h\rightarrow bs)$ are in the region (0.0015$\sim$0.006).

Through data analysing, we plot the relationship between $g_{YX}$ and Br$(h \rightarrow bs)$ in Fig.\ref{T1}(b).
 In the context of $U(1)_X$SSM, $g_{YX}$ is the mixing gauge coupling constant beyond MSSM,
 which affects the strength of the coupling vertices. The study of $g_{YX}$ is of interest.
 To investigate its impact on the  Br$(h \rightarrow bs)$, we vary $g_{YX}$ within the range of (0.1$-$0.24).
 and the values also increase when $g_{YX}$ turns large.
 The characteristics of Fig.\ref{T1}(a) are similar as those of Fig.\ref{T1}(b).
 That is to say, $g_X$ and $g_{YX}$ have the similar effects to a certain extent.

 $\lambda_H$ comes from the term $\lambda_HSH_uH_d$ in the superpotential.
 The mass matrixes of several
 particles(chargino, neutralino, down-squark, up-squark, neutral Higgs, charged Higgs)
 all have the important parameter $\lambda_H$, which possibly produces
 complex effects on the numerical results. The contribution from each diagram in the Fig.\ref{N1} is influenced by $\lambda_H$.
  Fig.$\ref{T1}$ (c) represents the relationship between $\lambda_H$ and Br$(h \rightarrow bs)$. Within the $\lambda_H$ range of 0.08 to 0.19, the corresponding values of the lines decrease as $\lambda_H$ increases.
  This characteristic should be the result of competition of different contributions from the one-loop diagrams. In the whole, the value of Br$(h \rightarrow bs)$ falls within the range of $(1\sim6) \times 10^{-3}$ from the three diagrams in the Fig.\ref{T1}.

\subsection{two-dimensional scatter plot(in the $U(1)_XSSM$)}
 To better study the influence of parameters, we draw some multidimensional scatter plots based on $\chi^2$ and select values that fall within  $3\sigma$ range of $\chi^2$, ensuring a more significant correlation between the variables and Br$(h \rightarrow bs)$. We utilize the simplified expression of $\chi^2$ as
\begin{eqnarray}
&&\chi^2 = \sum_i(\frac{\mu^{th}_i - \mu^{exp}_i}{\delta_i})^2.\label{th}
\end{eqnarray}
In Eq.(\ref{th}), $\mu^{th}_i$ signifies the theoretical value for the corresponding procedure derived within $U(1)_X$SSM. The experimental data is denoted as $\mu^{exp}_i$, while $\delta_i$ represents the error encompassing both statistical and systematic components.

The specific expression of $\chi^2$ is presented as
\begin{eqnarray}
&&\chi^2 = \Big(\frac{m^{th}_{h} - m^{exp}_{h}}{\delta_{m_{h}}}\Big)^2 + \Big(\frac{\mu^{th}_{\gamma\gamma} - \mu^{exp}_{\gamma\gamma}}{\delta_{{\gamma\gamma}}}\Big)^2 + \Big(\frac{\mu^{th}_{ZZ} - \mu^{exp}_{ZZ}}{\delta_{{ZZ}}}\Big)^2\nonumber\\&&
+ \Big(\frac{\mu^{th}_{WW} - \mu^{exp}_{WW}}{\delta_{{WW}}}\Big)^2
+ \Big(\frac{\mu^{th}_{b\bar b} - \mu^{exp}_{b\bar b}}{\delta_{{b\bar b}}}\Big)^2 + \Big(\frac{\mu^{th}_{\tau\bar\tau} - \mu^{exp}_{\tau\bar\tau}}{\delta_{{\tau\bar\tau}}}\Big)^2 + \Big(\frac{\Delta a^{th}_{\mu} - \Delta a_{\mu}}{\delta_{\Delta a_\mu}}\Big)^2.
\end{eqnarray}
The averaged values of the experimental data are derived from the updated PDG [20], $m^{exp}_{h}$ = 125.25 $\pm$ 0.17 GeV, $\mu^{exp}_{\gamma\gamma}$ = 1.10 $\pm$ 0.07, $\mu^{exp}_{ZZ}$ = 1.01 $\pm$ 0.07, $\mu^{exp}_{WW}$ = 1.19 $\pm$ 0.12, $\mu^{exp}_{b\bar b}$ = 0.98 $\pm$ 0.12, $\mu^{exp}_{\tau\bar\tau}$ = 1.15 $\pm$ 0.15. $\Delta a_\mu$ = (2.50 $\pm$ 0.48) $\times$ $10^{-9}$ is obtained from the latest work of muon g-2 [30-31].

\begin{table*}
\caption{Scanning parameters for Fig.\ref{T0}, Fig.\ref{T2}, Fig.\ref{T3} and Fig.\ref{T4}.}
\begin{tabular*}{\textwidth}{@{\extracolsep{\fill}}|l|l|l|l|l|l|l|l|l|@{}}
\hline
Parameters&$g_X$&$g_{YX}$&$\tan\beta$&$M_1/~{\rm TeV}$&$M_2/~{\rm TeV}$&$M_{BL}/~{\rm TeV}$&$M_{BB'}/~{\rm TeV}$ &$M_{\tilde{d}}^2/~{\rm TeV}^2$\\
\hline
Min&0.3&0.01&8&0.2&1&0.1&0.1&4~~\\
\hline
Max&0.4&0.3&35&1&2&1&1&16~~\\
\hline
\end{tabular*}
\label{B2}
\end{table*}

\begin{figure}[ht]
\setlength{\unitlength}{5mm}
\centering
\includegraphics[width=2.5in]{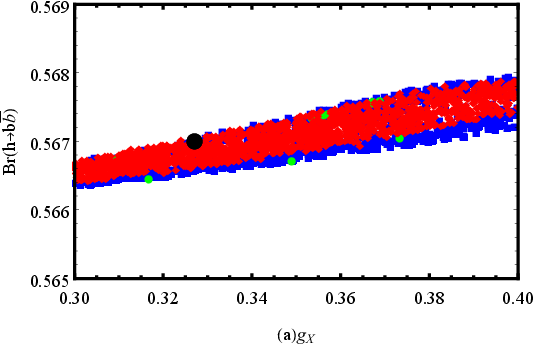}
\setlength{\unitlength}{5mm}
\centering
\includegraphics[width=2.5in]{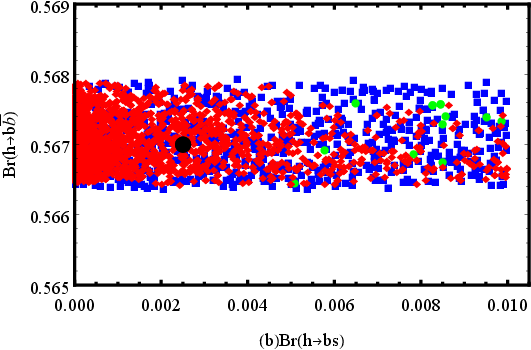}
\setlength{\unitlength}{5mm}
\caption{(a) Br$(h \rightarrow b\bar b)$ versus $g_X$; (b) Br$(h \rightarrow b\bar b)$ versus Br$(h \rightarrow bs)$. $\textcolor{red}{\blacklozenge}$ is a value located within the $1\sigma$ range of $\chi^2$ in the domain of $\chi^2 \leq$ 14.41. $\textcolor{blue}{\blacksquare}$ is a value situated within the $(1\sim2)\sigma$ range of $\chi^2$, where 14.41 $\leq \chi^2 \leq$ 20.89. $\textcolor{green}{\bullet}$ is a value situated within the $(2\sim3)\sigma$ range of $\chi^2$, where again 20.89 $\leq \chi^2 \leq$ 28.65. $\bullet$ represents the value corresponding to the optimal point.} {\label {T0}}
\end{figure}

\begin{figure}[ht]
\setlength{\unitlength}{5mm}
\centering
\includegraphics[width=2.5in]{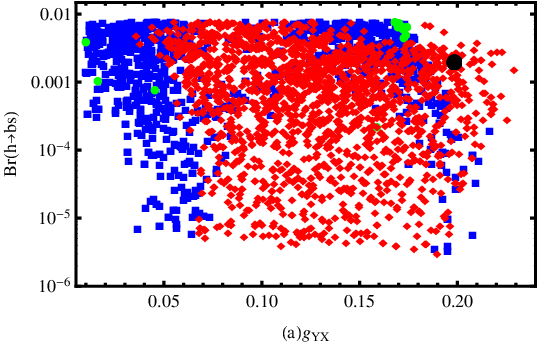}
\setlength{\unitlength}{5mm}
\centering
\includegraphics[width=2.5in]{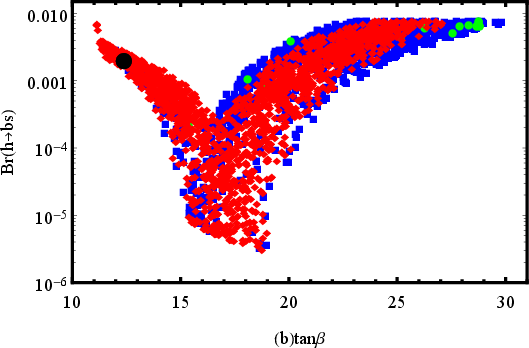}
\setlength{\unitlength}{5mm}
\caption{(a) The $\chi^2$ plot in the $g_{YX}-$Br$(h \rightarrow bs)$ plane; (b) The $\chi^2$ plot in the $\tan\beta-$Br$(h \rightarrow bs)$ plane. it has same color code as Fig.\ref {T0}.} {\label {T2}}
\end{figure}

\begin{figure}[ht]
\includegraphics[width=2.5in]{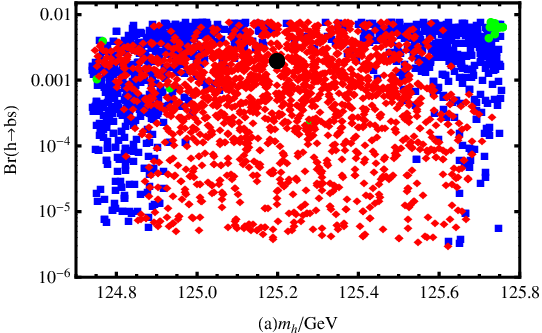}
\setlength{\unitlength}{5mm}
\centering
\includegraphics[width=2.5in]{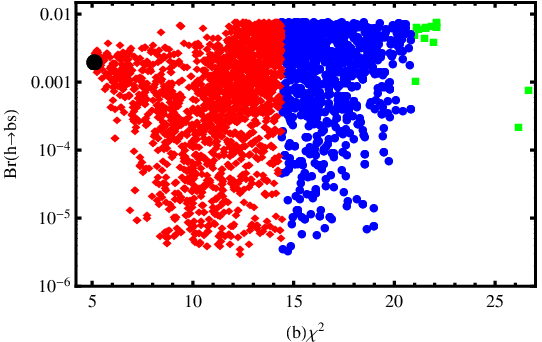}
\caption{(a) The $\chi^2$ plot in the $m_h-$Br$(h \rightarrow bs)$ plane; (b) The $\chi^2$ plot in the $\chi^2-$Br$(h \rightarrow bs)$ plane. it has same color code as Fig.\ref {T0}.} {\label {T3}}
\end{figure}

We randomly scan eight parameters and present them in Table $\ref{B2}$. Fig.$\ref{T0}$, Fig.$\ref{T2}$, Fig.$\ref{T3}$ and Fig.$\ref{T4}$ are plotted using the parameters present in Table $\ref{B2}$. $\textcolor{red}{\blacklozenge}$ is a value located within the $1\sigma$ range of $\chi^2$ in the domain of $\chi^2 \leq$ 14.41. $\textcolor{blue}{\blacksquare}$ is a value situated within the $(1\sim2)\sigma$ range of $\chi^2$, where 14.41 $\leq \chi^2 \leq$ 20.89. $\textcolor{green}{\bullet}$ is a value situated within the $(2\sim3)\sigma$ range of $\chi^2$, where again 20.89 $\leq \chi^2 \leq$ 28.65. $\bullet$ represents the value corresponding to the optimal point.

We chose two parameters to plot Br$(h \rightarrow b\bar b)$, which are $g_X$ and Br$(h \rightarrow bs)$. In Fig.$\ref{T0}$ (a), when $g_X$ increases, Br$(h \rightarrow b\bar b)$ has an upward trend, and remain in the $1\sigma$ range of SM prediction; In Fig.$\ref{T0}$ (b), most of the points are in the order of $10^{-3}$ of Br$(h \rightarrow bs)$, and the points of Br$(h \rightarrow b\bar b)$ are in the range of $(0.565 - 0.568)$.

Subsequently, from eight parameters, we select two sensitive parameters $g_{YX}$ and $\tan\beta$ to show the results, and plot the graphs respectively using $g_{YX}$ and $\tan\beta$ as the horizontal axes and Br$(h \rightarrow bs)$ as the vertical axis in Fig.$\ref{T2}$. These two parameters exhibit significant influence on Br$(h \rightarrow bs)$. As seen in Fig.$\ref{T2}$(a), most of the points are distributed within the range of 0.05 $\leq$ $g_{YX}$ $\leq$ 0.20 and $10^{-4}$ $\leq$ Br$(h \rightarrow bs)$ $\leq$ $10^{-2}$. When $g_{YX}$ $\geq$ 0.2 and Br$(h \rightarrow bs)$ $\leq$ $10^{-4}$, the density of points gradually decreases. In Fig.$\ref{T2}$(b), the majority of points are situated within the range of 11 $\leq$ $\tan\beta$ $\leq$ 29 and 5 $\times$ $10^{-4}$ $\leq$ Br$(h \rightarrow bs)$ $\leq$ 0.01. When 15 $\leq$ $\tan\beta$ $\leq$ 20 and Br$(h \rightarrow bs)$ $\leq$ 5 $\times$ $10^{-4}$, the distribution of points becomes less dense. It is evident that when $g_{YX}$ $\approx$ 0.20 and $\tan\beta$ $\approx$ 12, $\chi^2$ reaches its optimal value, resulting in Br$(h \rightarrow bs)$ of $2.5 \times 10^{-3}$.

Here, in Fig.$\ref{T3}$(a), we present a scatter plot of $m_h$ with Br$(h \rightarrow bs)$. It can be observed that the optimal point of $\chi^2$ is located near $m_h$ = 125.2 GeV, which aligns well with the mass of the CP-even Higgs. As the mass of $m_h$ increases or decreases from 125.2 GeV within the 3$\sigma$ range, the density of points gradually enhances, and the points also cluster around the Br$(h \rightarrow bs)$ at the order of $10^{-3}$. Fig.$\ref{T3}$(b) depicts a scatter plot of $\chi^2$ with Br$(h \rightarrow bs)$, clearly illustrating the distribution of Br$(h\rightarrow bs)$ as $\chi^2$ varies.

\subsection{the Significant Impact of $a_\mu$ on Numerical Calculations(in the $U(1)_XSSM$)}
Next, we investigate the relation of muon magnetic dipole moment (MDM) with the flavor violation of $h \rightarrow bs$. The E989 collaboration at Fermilab \cite{MDM1} report the latest experimental data as $a_{\mu}^{FNAL}=116592055(24)\times 10^{-11}$. The new averaged experiment value of muon MDM is $a^{exp}_{\mu}=116592059(22)\times 10^{-11}$.
Therefore, there is a deviation of 5.0$\sigma$ between the experiment and SM expectation ($\Delta a_\mu=a^{exp}_\mu-a^{SM}_\mu=249(48)\times 10^{-11}$)\cite{MDM2}.

\begin{figure}[ht]
\setlength{\unitlength}{5mm}
\centering
\includegraphics[width=2.5in]{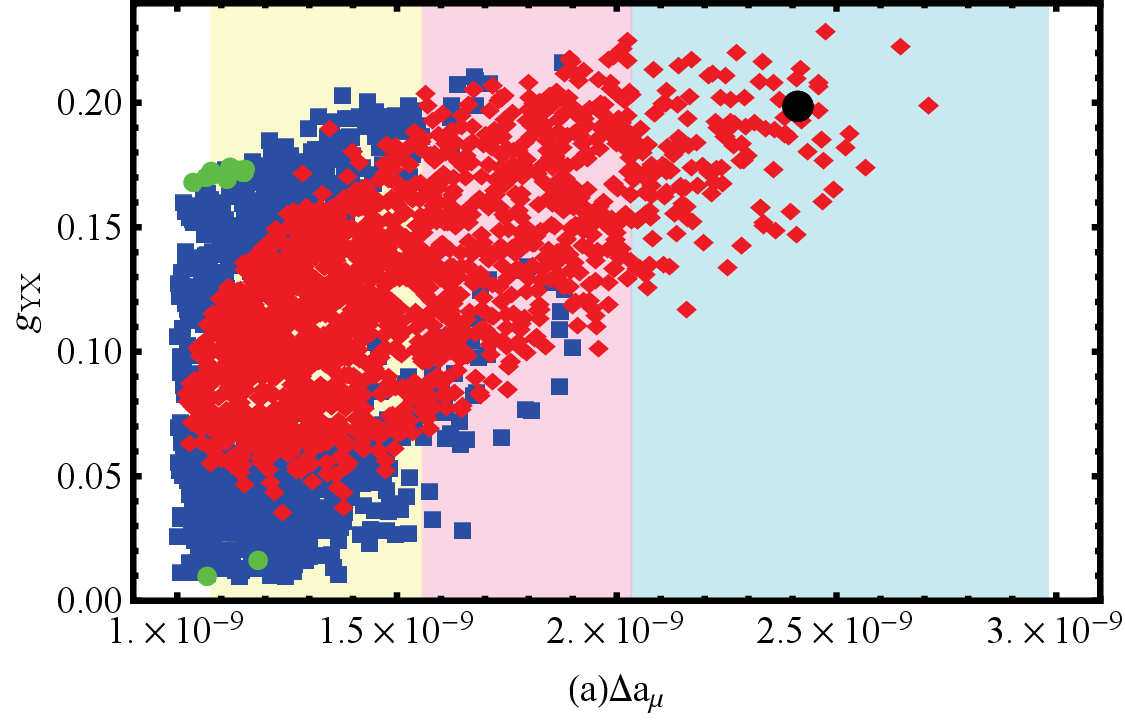}
\setlength{\unitlength}{5mm}
\centering
\includegraphics[width=2.5in]{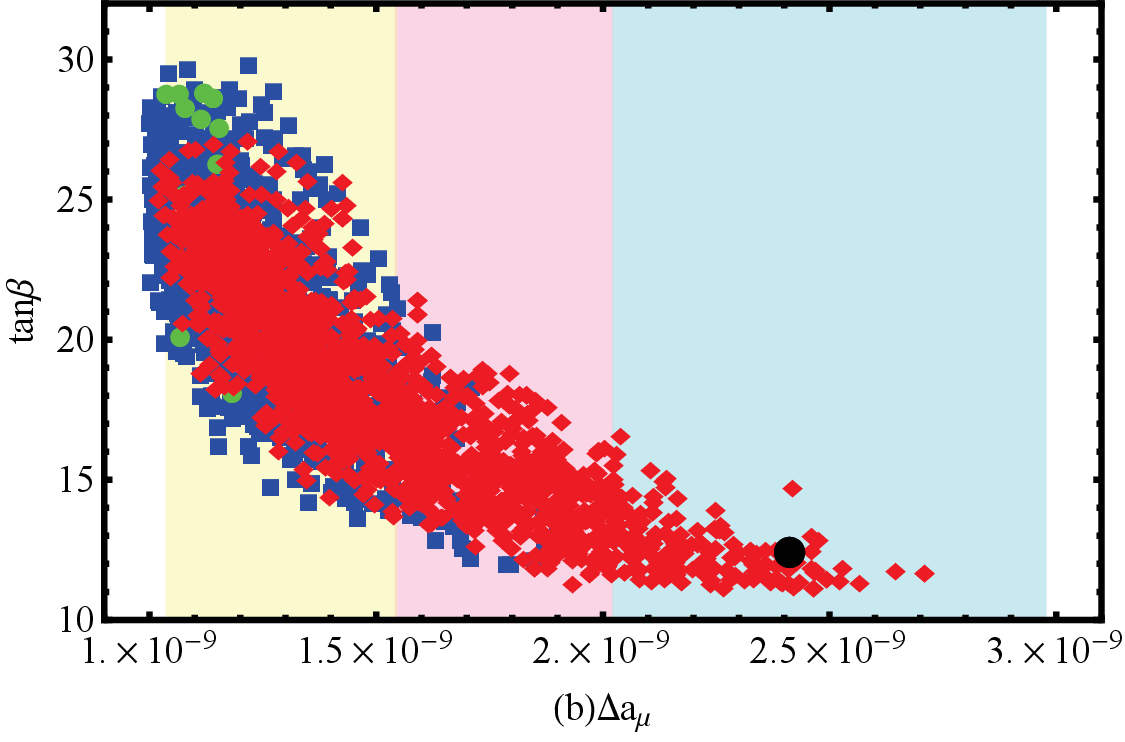}
\setlength{\unitlength}{5mm}
\centering\nonumber\\
\includegraphics[width=2.5in]{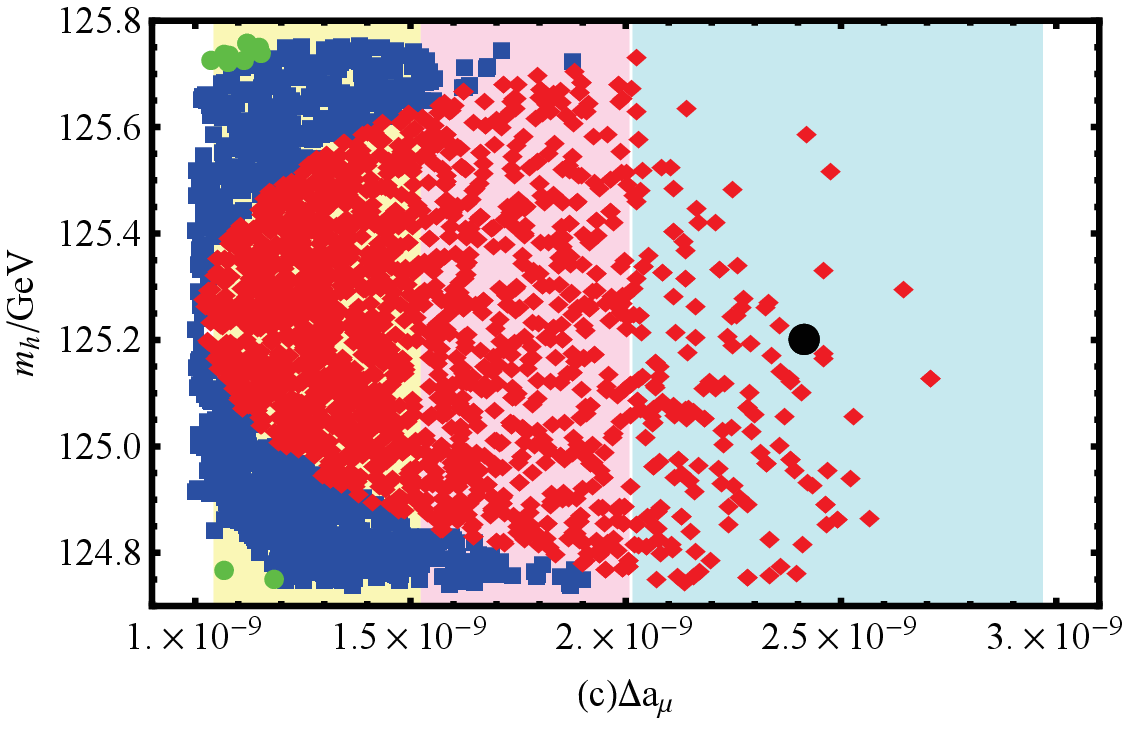}
\setlength{\unitlength}{5mm}
\centering
\includegraphics[width=2.5in]{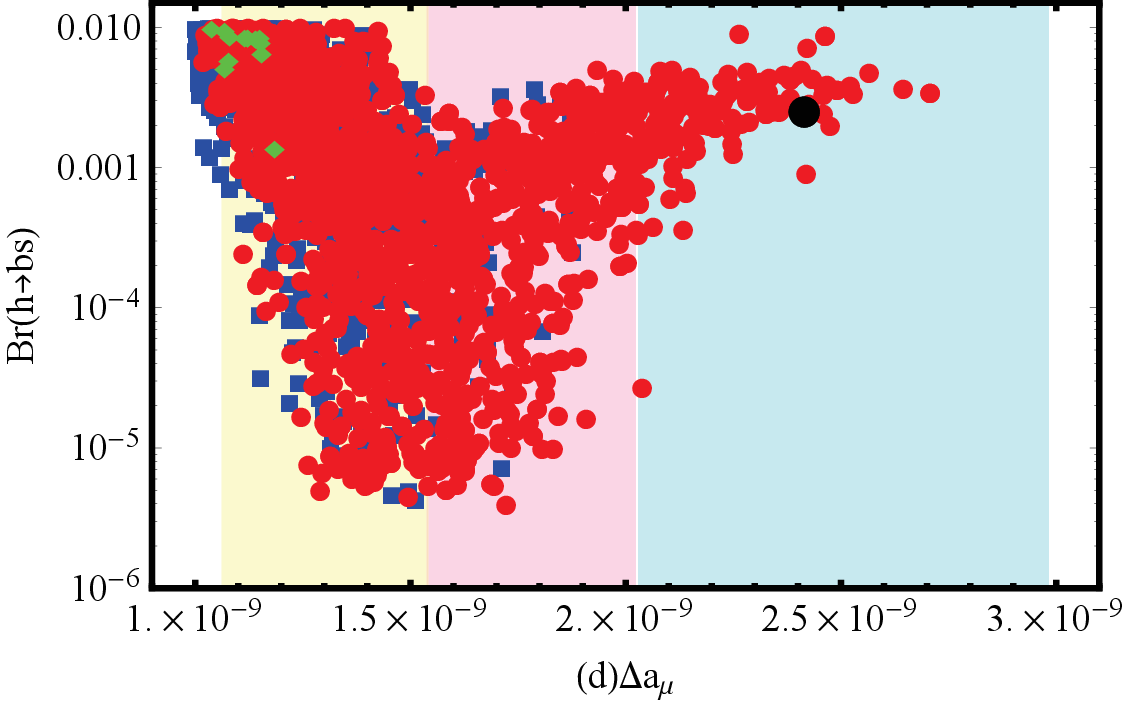}
\setlength{\unitlength}{5mm}
\centering
\includegraphics[width=2.5in]{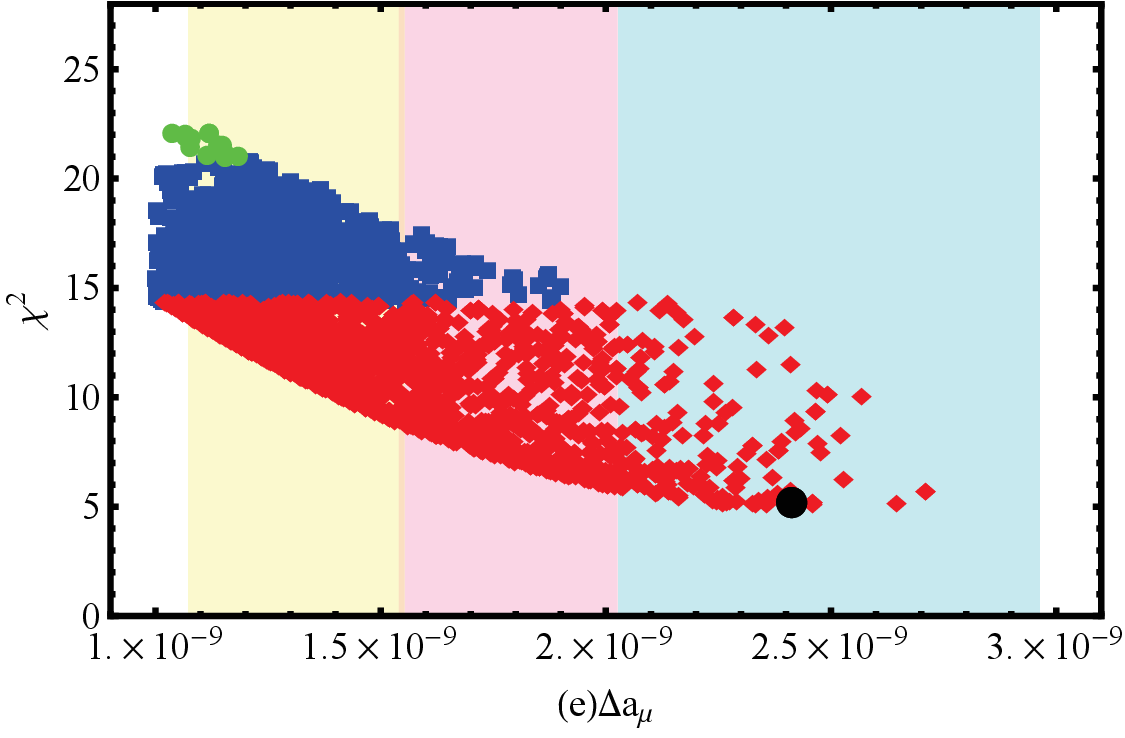}
\caption{The relation of $\Delta a_\mu$ with other parameters. It has same color code as Fig.\ref {T0}.} {\label {T4}}
\end{figure}

In Fig.$\ref{T4}$, we present a series of scatter plots to illustrate the correlations between $\tan\beta$, $m_h$, $g_{YX}$, Br$(h \rightarrow bs)$, $\chi^2$, and $\Delta a_\mu$. To clarify, the blue portion represents the $1\sigma (2.02 \times 10^{-9} \leq \Delta a_\mu \leq 2.98 \times 10^{-9})$ range of $\Delta a_\mu$, the pink portion represents the $2\sigma (1.54 \times 10^{-9} \leq \Delta a_\mu \leq 2.02 \times 10^{-9})$ range, and the yellow portion represents the $3\sigma (1.06 \times 10^{-9} \leq \Delta a_\mu \leq 1.54 \times 10^{-9})$ range. We can easily observe that the numerical results of $\chi^2$ are all within the $3\sigma$ experimental limit of $\Delta a_\mu$, and the $\chi^2$ optimal point is located at the 1$\sigma (2.02 \times 10^{-9} \leq \Delta a_\mu \leq 2.50 \times 10^{-9})$ position of $\Delta a_\mu$. In the area $\Delta a_\mu \in [(2.50 + 0.48), (2.50 + 1.44)] \times 10^{-9}$, there are no points in our used parameter space. So we do not show them. These diagrams indicate that when we take the optimal value of $\chi^2$, the numerical results of $\Delta a_\mu$ can effectively compensate for the deviation between the experimental data and the SM predictions.

The Fig.\ref{T4}(a) is plotted in the plane of $\Delta a_\mu$ versus $g_{YX}$. In the Fig.\ref{T4}(a), there are a lot more $\textcolor{red}{\blacklozenge}$ and $\textcolor{blue}{\blacksquare}$ than $\textcolor{green}{\bullet}$,
 and their distribution is basically symmetrical. More points are located in the middle and left region.
In the top right corner, there are $\textcolor{red}{\blacklozenge}$ and $\textcolor{black}{\bullet}$.
It implies that large $g_{YX}$ leads to large $\Delta a_\mu$.
The Fig.\ref{T4}(b) shows the relation between $\Delta a_\mu$ and $\tan\beta$,
where the outline of the points gradually decreases with enlarging $\Delta a_\mu$.
In our used parameter space, $\Delta a_\mu$ is the decreasing function of $\tan\beta$.
The diagram of $m_h$ versus $\Delta a_\mu$ is shown by the Fig.\ref{T4}(c).
The image of the points is almost symmetrical with respect to the line $m_h=125.25{\rm GeV}$.
The interior is $\textcolor{red}{\blacklozenge}$, and outwards are $\textcolor{blue}{\blacksquare}$ and \textcolor{green}{ $\bullet$} in turn.
The relationship between $\Delta a_\mu$ and Br$(h\rightarrow bs)$ is worth discussing,
which is expressed by Fig.\ref{T4}(d). With the scanned parameters,
when $\Delta a_\mu$ is larger than $1.5\times 10^{-9}$, $\Delta a_\mu$ and Br$(h\rightarrow bs)$ present the relationship of the increasing function.
Both $\Delta a_{\mu}$ and Br$(h \rightarrow bs)$  are increasing functions with the enlarging $g_{YX}$.
$\Delta a_{\mu}$ is the decreasing function of $\tan\beta$ in the used parameter space. Br$(h \rightarrow bs)$  decreases first and then increases as $g_{YX}$ increases.
Therefore, $\Delta a_{\mu}$ and Br$(h \rightarrow bs)$ show similar trend during $\tan\beta$ region where Br$(h \rightarrow bs)$  is the decreasing function of $\tan\beta$.
In the whole, $\Delta a_\mu$ and Br$(h\rightarrow bs)$ show the same trend, when they vary with $g_{YX}$ and $\tan\beta$.
Therefore, as $\Delta a_\mu$ improves, the prediction for the Br$(h\rightarrow bs)$ increases.
In the Fig.\ref{T4}(e), when $\Delta a_\mu$ turns large, $\chi^2$ shrinks gradually.

The parameters
$\tan\beta$, $m_h$, and $g_{YX}$ are still significant factors that have a considerable impact on Br$(h\rightarrow bs)$. Compared to the previous section, after adding restrictions of $\Delta a_\mu$, we obtain more precise parameter spaces.

\section{Conclusion}
In the $U(1)_X$SSM, we analyze the phenomenology of quark flavor-changing decays of the Higgs boson to a strange-bottom quark pair. We assume that the lightest CP-even Higgs boson, denoted as $h$, is the SM-like Higgs boson discovered by the LHC. A significant focus of this paper is to analyze the $U(1)_X$SSM contributions to $h\rightarrow bs$.

From the perspective of the magnitude of Br$(h \rightarrow bs)$ and data analysis, the Br$(h \rightarrow bs)$ can reach the order of $10^{-3}$, which is four orders of magnitude larger than the SM prediction. We consider the Feynman diagrams for the $h \rightarrow bs$ process and perform extensive calculations, plotting linear graphs of various parameters versus Br$(h \rightarrow bs)$. To enhance the fitting degree of the numerical analysis, we include $\chi^2$ and conduct a broad scan of the parameters. Through numerical analysis, we find that a subset of parameters exert a significant influence on the results. $M_1$, $M_2$, $M_{BL}$, $M_{BB'}$, and $M^2_{\tilde{D}}$ are the unsensitive parameters, while $g_X$, $g_{YX}$, $\tan\beta$, and $\lambda_H$ are the sensitive parameters. Finally, we analyze the relationship between $\Delta a_\mu$ and other sensitive parameters, as well as the correlation between $\Delta a_\mu$ and Br$(h \rightarrow bs)$. Through comprehensive research, we have demonstrated that all parameters in our work satisfy the constraints from $h \rightarrow bs$ process. However, such decays remain undetectable at the LHC\cite{Finally1,Finally2}. It may be detected at ILC in the future. We believe that our findings may serve as a valuable resource for those interesting decays within alternative frameworks of new physics.

\begin{acknowledgments}
This work is supported by National Natural Science Foundation of China (NNSFC)(No.12075074),
Natural Science Foundation of Hebei Province(A2020201002, A2023201040, A2022201022, A2022201017, A2023201041),
Natural Science Foundation of Hebei Education Department (QN2022173),
Post-graduate's Innovation Fund Project of Hebei University (HBU2024SS042, Study the Higgs boson decay $h\rightarrow bs$ in the $U(1)_X$SSM).
\end{acknowledgments}

\appendix
\section{Used mass matrices and coupling vertices in $U(1)_X$SSM}

At the tree level, the mass-squared matrix for the CP-even Higgs
($\phi_d,\phi_u,\phi_\eta,\bar\phi_\eta,\phi_s$) is as follows
\begin{eqnarray}
&&M^2_h = \left(\begin{array}{ccccc}m_{\phi_d\phi_d}&m_{\phi_u\phi_d}&m_{\phi_\eta\phi_d}
&m_{\phi_{\bar\eta}\phi_d}&m_{\phi_s\phi_d}\\
m_{\phi_d\phi_u}&m_{\phi_u\phi_u}&m_{\phi_\eta\phi_u}
&m_{\phi_{\bar\eta}\phi_u}&m_{\phi_s\phi_u}\\
m_{\phi_d\phi_\eta}&m_{\phi_u\phi_\eta}&m_{\phi_\eta\phi_\eta}
&m_{\phi_{\bar\eta}\phi_\eta}&m_{\phi_s\phi_\eta}\\
m_{\phi_d\phi_{\bar\eta}}&m_{\phi_u\phi_{\bar\eta}}&m_{\phi_\eta\phi_{\bar\eta}}
&m_{\phi_{\bar\eta}\phi_{\bar\eta}}&m_{\phi_s\phi_{\bar\eta}}\\
m_{\phi_d\phi_s}&m_{\phi_u\phi_s}&m_{\phi_\eta\phi_s}
&m_{\phi_{\bar\eta}\phi_s}&m_{\phi_s\phi_s}\end{array}\right),\indent
\end{eqnarray}
\begin{eqnarray}
&&m_{\phi_d\phi_d} = m_{H_d}^2 + \mu^2 + \frac{1}{8}\Big([g_1^2 + (g_X + g_{YX})^2 + g_2^2](3v_d^2 - v_u^2)\nonumber\\&&\hspace{1.8cm}+2(g_{YX}g_X + g_X^2)(v_\eta^2 - v_{\bar\eta}^2)\Big) + \sqrt{2}v_S\mu\lambda_H + \frac{1}{2}(v_u^2 + v_S^2)\lambda_H^2,\\
&&m_{\phi_d\phi_u} = -\frac{1}{4}\Big(g_2^2 + (g_{YX} + g_X)^2 + g_1^2\Big)v_dv_u + \lambda_H^2v_dv_u - \lambda_H l_W\nonumber\\&&\hspace{1.8cm} -\frac{1}{2}\lambda_H v_\eta v_{\bar\eta}\lambda_C + v_S^2\kappa - B_\mu - \sqrt{2}v_S(\frac{1}{2}T_{\lambda_H}+M_S\lambda_H),\\
&&m_{\phi_u\phi_u} = m_{H_u}^2 + \mu^2 + \frac{1}{8}\Big\{\Big([g_1^2 + (g_X + g_{YX})^2 + g_2^2]\Big)(3v_u^2 - v_d^2)\nonumber\\&&\hspace{1.8cm}+2(g_{YX}g_X + g_X^2)( v_{\bar\eta}^2 - v_\eta^2 )\Big\} + \sqrt{2}v_S\mu\lambda_H + \frac{1}{2}(v_d^2 + v_S^2)\lambda_H^2,\\
&&m_{\phi_d\phi_\eta} = \frac{1}{2}g_X(g_{YX} + g_X)v_dv_\eta -\frac{1}{2}v_uv_{\bar\eta}\lambda_H\lambda_C,\\
&&m_{\phi_u\phi_\eta} = -\frac{1}{2}g_X(g_{YX} + g_X)v_dv_\eta -\frac{1}{2}v_dv_{\bar\eta}\lambda_H\lambda_C,\\
&&m_{\phi_\eta\phi_\eta} = m_\eta^2 + \frac{1}{4}\Big([(g_{YX}g_X + g_X^2)](v_d^2 - v_u^2)+2g_X^2(3v_\eta^2-v_{\bar\eta}^2)\Big) + \frac{\lambda_C^2}{2}(v_{\bar\eta}^2 + v_S^2),\\
&&m_{\phi_d\phi_{\bar\eta}} = -\frac{1}{2}g_X(g_{YX} + g_X)v_dv_\eta -\frac{1}{2}v_uv_{\bar\eta}\lambda_H\lambda_C,\\
&&m_{\phi_u\phi_{\bar\eta}} = \frac{1}{2}g_X(g_{YX} + g_X)v_dv_\eta -\frac{1}{2}v_dv_{\bar\eta}\lambda_H\lambda_C,\\
&&m_{\phi_\eta\phi_{\bar\eta}} = ({\lambda}_C^2 - g_X^2)v_\eta v_{\bar\eta} + \frac{\lambda_C}{2}(2l_W - \lambda_Hv_dv_u) + \frac{v_S}{\sqrt{2}}(2M_S\lambda_C + T_{\lambda_C}) + \frac{v_S^2}{2}\lambda_C\kappa,\\
&&m_{\phi_{\bar\eta}\phi_{\bar\eta}} = m_{\bar\eta}^2 + \frac{1}{4}\Big((g_{YX}g_X + g_X^2)(v_u^2 - v_d^2)+2g_X^2(3v_{\bar\eta}^2-v_\eta^2)\Big) + \frac{\lambda_C^2}{2}(v_\eta^2 + v_S^2),\\
&&m_{\phi_d\phi_s} = \Big(\lambda_Hv_dv_S + \sqrt{2}v_d\mu - v_u(\kappa v_S + \sqrt{2}M_S)\Big)\lambda_H - \frac{1}{\sqrt{2}}v_uT_{\lambda_H},\\
&&m_{\phi_u\phi_s} = \Big(\lambda_Hv_uv_S + \sqrt{2}v_u\mu - v_d(\kappa v_S + \sqrt{2}M_S)\Big)\lambda_H - \frac{1}{\sqrt{2}}v_dT_{\lambda_H},\\
&&m_{\phi_\eta\phi_s} = \Big(\lambda_Cv_\eta v_S + v_{\bar\eta}(\kappa v_S + \sqrt{2}M_S)\Big)\lambda_C + \frac{1}{\sqrt{2}}v_{\bar\eta}T_{\lambda_C},\\
&&m_{\phi_{\bar\eta}\phi_s} = \Big(\lambda_Cv_{\bar\eta} v_S + v_\eta(\kappa v_S + \sqrt{2}M_S)\Big)\lambda_C + \frac{1}{\sqrt{2}}v_\eta T_{\lambda_C},\\
&&m_{\phi_s\phi_s} = m_S^2 + \Big(2l_W + 3v_S(\kappa v_S + 2\sqrt{2}M_S) + \lambda_Cv_\eta v_{\bar\eta} - \lambda_Hv_dv_u \Big)\kappa + 2B_S\nonumber\\&&\hspace{1.8cm} +\frac{1}{2}\lambda_C^2\xi^2 + \frac{1}{2}\lambda_H^2v^2 + 4M_S^2 + \sqrt{2}v_ST_\kappa.
\end{eqnarray}
This matrix is diagonalized by $Z^H$
\begin{equation}
Z^HM_h^2Z^{H,\dagger} = M^{dia}_{2,h}.
\end{equation}
The mass matrix for charginos in the basis: ($\tilde{W}^-$,$\tilde{H}_d^-$), ($\tilde{W}^+$,$\tilde{H}_u^+$)
\begin{eqnarray}
M_{{\chi}^\pm}=
\left({\begin{array}{*{20}{c}}
M_2 & \frac{1}{\sqrt{2}}g_2v_u \\
\frac{1}{\sqrt{2}}g_2v_d & \frac{1}{\sqrt{2}}\lambda_Hv_S+\mu \\
\end{array}}
\right),
\end{eqnarray}
The matrix is diagonalized by U and V
\begin{eqnarray}
U^*M_{{\chi}^\pm} V^\dag = M_{{\chi}^\pm}^{dia}.
\end{eqnarray}
The mass squared matrix for up-squarks ($\tilde{u}^0_L , \tilde{u}^0_R$) reads
\begin{eqnarray}
&&M^2_{\tilde{U}} = \left(\begin{array}{cc}m_{\tilde{u}^0_L{\tilde{u}^{0,*}_L}}
&m^\dagger_{\tilde{u}^0_R{\tilde{u}^{0,*}_L}}\\
m_{\tilde{u}^0_L{\tilde{u}^{0,*}_R}}&m_{\tilde{u}^0_R{\tilde{u}^{0,*}_R}}\end{array}\right),\indent
\end{eqnarray}
\begin{eqnarray}
&&m_{\tilde{u}^0_L{\tilde{u}^{0,*}_L}} = \frac{1}{24}\Big(3g_2^2(-v_u^2 + v_d^2) + (g_1^2 + g_{YX}^2)(-v_d^2 + v_u^2)\nonumber\\&&\hspace{1.8cm}
 + g_{YX}g_X(2v_{\bar\eta}^2 - 2v_\eta^2 - v_d^2 + v_u^2)\Big) +\frac{1}{2}(2m_{\tilde Q}^2 + v_u^2Y^\dagger_uY_u),\\
&&m_{\tilde{u}^0_L{\tilde{u}^{0,*}_R}} = -\frac{1}{2}\Big(\sqrt{2}(v_dY_u\mu^* - v_uT_u) + v_dv_SY_u\lambda^*_H\Big),\\
&&m_{\tilde{u}^0_R{\tilde{u}^{0,*}_R}} = \frac{1}{24}\Big(4(g_1^2 + g_{YX}^2)(-v_u^2 + v_d^2) + g_{YX}g_X(7v_d^2 - 7v_u^2 - 8v_{\bar\eta}^2 + 8v_\eta^2)\nonumber\\&&\hspace{1.8cm} + 3g_X^2(-2v_{\bar\eta}^2 + 2v_\eta^2 - v_u^2 +v_d^2)\Big) + \frac{1}{2}(2m_{\tilde U}^2 + v_u^2Y_uY^\dagger_u).
\end{eqnarray}
The mass squared matrix for down-squarks ($\tilde{d}^0_L , \tilde{d}^0_R$) reads
\begin{eqnarray}
&&M^2_{\tilde{D}} = \left(\begin{array}{cc}m_{\tilde{d}^0_L{\tilde{d}^{0,*}_L}}
&m^\dagger_{\tilde{d}^0_R{\tilde{d}^{0,*}_L}}\\
m_{\tilde{d}^0_L{\tilde{d}^{0,*}_R}}&m_{\tilde{d}^0_R{\tilde{d}^{0,*}_R}}\end{array}\right),\indent
\end{eqnarray}
\begin{eqnarray}
&&m_{\tilde{d}^0_L{\tilde{d}^{0,*}_L}} = \frac{1}{24}\Big((3g_2^2 + g_1^2 + g_{YX}^2)(-v_d^2 + v_u^2) + g_{YX}g_X(2v_{\bar\eta}^2 - 2v_\eta^2 - v_d^2 + v_u^2)\Big)\nonumber\\&&\hspace{1.8cm} +\frac{1}{2}(2m_{\tilde{Q}}^2 + v_d^2Y^\dagger_dY_d),\\
&&m_{\tilde{d}^0_L{\tilde{d}^{0,*}_R}} = -\frac{1}{2}\Big(\sqrt{2}(-v_dT_d + v_uY_d\mu^*) + v_uv_SY_d\lambda^*_H\Big),\\
&&m_{\tilde{d}^0_R{\tilde{d}^{0,*}_R}} = \frac{1}{24}\Big(2(g_1^2 + g_{YX}^2)(-v_u^2 + v_d^2) + g_{YX}g_X(-4v_{\bar\eta}^2 + 4v_\eta^2 + 5v_d^2 - 5v_u^2)\nonumber\\&&\hspace{1.8cm} + 3g_X^2(-2v_{\bar\eta}^2 + 2v_\eta^2 - v_u^2 +v_d^2)\Big) + \frac{1}{2}(2m_{\tilde D}^2 + v_d^2Y_dY^\dagger_d).
\end{eqnarray}
The mass matrix for neutralino in the basis $(\lambda_{\tilde{B}},\tilde{W}^0,\tilde{H}_d^0,\tilde{H}_u^0,\lambda_{\tilde{X}},\tilde{\eta},\tilde{\bar\eta},\tilde{s})$ is
\begin{eqnarray}
M_{{\chi}^0}=
\left({\begin{array}{*{20}{c}}
M_1 & 0 & -\frac{g_1}{2}v_d & \frac{g_1}{2}v_u & M_{{BB}^{\prime}} & 0 & 0 & 0 \\
0 & M_2 & \frac{g_2}{2}v_d & -\frac{g_2}{2}v_u & 0 & 0 & 0 & 0 \\
-\frac{g_1}{2}v_d & \frac{g_2}{2}v_d & 0 & m_{{\tilde{H}_u^0}{\tilde{H}_d^0}} & m_{\lambda_{\bar{X}}\tilde{H}_d^0} & 0 & 0 & -\frac{\lambda_Hv_u}{\sqrt{2}} \\
\frac{g_1}{2}v_u & -\frac{g_2}{2}v_u & m_{{\tilde{H}_d^0}{\tilde{H}_u^0}} & 0 & m_{\lambda_{\bar{X}}{\tilde{H}_u^0}} & 0 & 0 & -\frac{\lambda_Hv_d}{\sqrt{2}} \\
M_{{BB}^\prime} & 0 & m_{\tilde{H}_d^0\lambda_{\bar{X}}} & m_{\tilde{H}_u^0\lambda_{\bar{X}}} & M_{BL} & -g_X{v_\eta} & g_Xv_{\bar{\eta}} & 0 \\
0 & 0 & 0 & 0 & -g_X{v_\eta} & 0 & \frac{1}{\sqrt{2}}\lambda_Cv_S & \frac{1}{\sqrt{2}}\lambda_Cv_{\bar{\eta}} \\
0 & 0 & 0 & 0 & g_Xv_{\bar{\eta}} & \frac{1}{\sqrt{2}}\lambda_Cv_S & 0 & \frac{1}{\sqrt{2}}\lambda_Cv_\eta \\
0 & 0 & -\frac{1}{\sqrt{2}}\lambda_Hv_u & -\frac{1}{\sqrt{2}}\lambda_Hv_d & 0 & \frac{1}{\sqrt{2}}\lambda_Cv_{\bar{\eta}} &  \frac{1}{\sqrt{2}}\lambda_Cv_\eta & m_{\tilde{s}\tilde{s}} \\
\end{array}}
\right),
\end{eqnarray}
\begin{eqnarray}
&&m_{{\tilde{H}_d^0}{\tilde{H}_u^0}}=-\frac{1}{\sqrt{2}}\lambda_Hv_S - \mu, ~~ m_{{\tilde{H}_d^0}\lambda_{\bar{X}}}=-\frac{1}{2}(g_{YX}+g_X){v_d},
\nonumber\\&&\ m_{\tilde{H}_u^0\lambda_{\bar{X}}}=\frac{1}{2}(g_{YX}+g_X)v_u, ~~ m_{\tilde{s}\tilde{s}}=2M_S+\sqrt{2}\kappa v_S.
\end{eqnarray}
The mass matrix for charged Higgs in the basis: ($H_d^{-}$,$H_u^{+,*}$), ($H_d^{-,*}$,$H_u^{+}$)
\begin{eqnarray}
M_{H^-}^2=
\left({\begin{array}{*{20}{c}}
m_{{H_d^{-}}H_d^{-,*}} & m_{H_u^{+,*}H_d^{-,*}}^{*} \\
m_{H_d^{-}H_u^{+}} & m_{H_u^{+,*}H_u^{+}} \\
\end{array}}
\right),
\end{eqnarray}
\begin{eqnarray}
&&m_{{H_d^{-}}H_d^{-,*}}=\frac{1}{8}\Big((g_2^2+g_X^2)v_d^2+(g_2^2-g_X^2)v_u^2+(g_1^2+g_{YX}^2)(v_d^2-v_u^2)-2g_X^2v_{\bar{\eta}}^2
\nonumber\\&&\hspace{1.8cm}+2[g_{YX}g_X(v_d^2+v_\eta^2-v_{\bar{\eta}}^2-v_u^2)+g_X^2v_\eta^2]\Big)\nonumber\\&&\hspace{1.8cm}
+\Big(\mid\mu\mid^2+\sqrt{2}v_S\Re(\mu\lambda_H^*)+\frac{1}{2}v_S^2\mid\lambda_H\mid^2 \Big),\\
&&m_{H_d^{-}H_u^{+}}=\frac{1}{2}\Big(2(\lambda_Hl_W^*+B_\mu)+\lambda_H(2\sqrt{2}v_S M_S^*-v_dv_u\lambda_H^*+v_\eta v_{\bar{\eta}}\lambda_C^*\nonumber\\&&\hspace{1.8cm}+\sqrt{2}v_S T_{\lambda_H})\Big)
+\frac{1}{4}g_2^2v_dv_u,\\
&&m_{H_u^{+,*}H_u^{+}}=\frac{1}{8}\Big((g_2^2-g_X^2)v_d^2+(g_2^2+g_X^2)v_u^2+(g_1^2+g_{YX}^2)(v_u^2-v_d^2)-2g_X^2v_\eta^2
\nonumber\\&&\hspace{1.8cm}+2[g_{YX}g_X(v_u^2+v_{\bar{\eta}}^2-v_d^2-v_\eta^2)+g_X^2v_{\bar{\eta}}^2]\Big)
\nonumber\\&&\hspace{1.8cm}+\frac{1}{2}\Big(2\mid\mu\mid^2+2\sqrt{2}v_S\Re(\mu\lambda_H^*)+v_S^2\mid\lambda_H\mid^2 \Big).
\end{eqnarray}

Here, we present some couplings required in this model
\begin{eqnarray}
&&\mathcal{L}_{{\chi}^+d{\tilde{u}^{*}}} = \sum_{a=1}^3\bar{\chi}_i^+\Big((-g_2V_{i1}^*Z^U_{ka} + V_{i2}^*Y_{u,a}Z^U_{k3+a})P_L+(Y^*_{d,a}Z^U_{ka}U_{i2})P_R\Big)d_{j\beta}\tilde{u}^{*}_{k\gamma}C_{ja},\indent\\&&
\mathcal{L}_{{\chi}^0d{\tilde{d}^{*}}} = -\frac{1}{6}\bar{\chi}^0_i\Big\{(\sqrt{2}g_{1}N^*_{i1}Z^D_{ka} - 3\sqrt{2}g_{2}N^*_{i2}Z^D_{ka} + \sqrt{2}g_{YX}N^*_{i5}Z^D_{ka} + 6N^*_{i3}Y_{d,a}Z^D_{k3+a})P_L + \indent\nonumber\\&&\Big(6Y^*_{d,a}Z^D_{kb}N_{i3} + \sqrt{2}Z^D_{k3+a}[2g_{1}N_{i1} + (2g_{YX} + 3g_X)N_{i5}]\Big)P_R\Big\} d_{j\beta}\tilde{d}^*_{k\gamma},\\&&
\mathcal{L}_{hH^-W^+} = \frac{1}{2}g_2h_i\Big(Z^H_{i1}Z^-_{j1} - Z^H_{i2}Z^-_{j2}\Big)\Big(-p_\mu^{H^-_j} + p_\mu^{h_i}\Big)H^-_jW^+_\mu,\\&&
\mathcal{L}_{hW^+W^-} =  \frac{1}{2}g_2^2h_i\Big(v_dZ^H_{i1} + v_uZ^H_{i2}\Big) g_{\sigma\mu}W{^+_\sigma}W{^-_\mu}.
\end{eqnarray}

Here $P_L=\frac{1-\gamma^5}{2}$, $P_R=\frac{1+\gamma^5}{2}$, $N$,  $Z^D$ and $Z^U$, are the diagonalizing matrices for $M_{{\chi}^0}$,  $M^2_{\tilde{D}}$ and $M^2_{\tilde{U}}$. They satisfy the relations $N^{T} M_{{\chi}^0} N=diag(m_{{\chi}^0_1},m_{{\chi}^0_2},...,m_{{\chi}^0_8})$
and  $Z_{\tilde{D}(\tilde{U})}^{\dagger} M^2_{\tilde{D}(\tilde{U})} Z_{\tilde{D}(\tilde{U})}=diag(m^2_{\tilde{d}(\tilde{u})_1},m^2_{\tilde{d}(\tilde{u})_2},...,m^2_{\tilde{d}(\tilde{u})_6})$, where $m_{{\chi}^0_i}(i=1,2,...,8)$ and $m^2_{\tilde{d}(\tilde{u})_i}(i=1,2,...,6)$ denote the
corresponding mass eigenvalues of $M_{{\chi}^0}$ and $M^2_{\tilde{D}(\tilde{U})}$.

\section{Feynman amplitude calculation process}
We initially address the resolution of the Feynman integral, utilizing the formula specified in Ref.\cite{Du4} for the integration of the denominator
\begin{eqnarray}
\frac{1}{ABC} = \int_{0}^{1}dx\int_{0}^{1}2ydy\frac{1}{[(Ax + B(1 - x))y + C(1 - y)]^3}.\label{x1}
\end{eqnarray}
Performing the integral calculation in this manner can significantly enhance the efficiency of numerical computations in our work. Based on Eq.(\ref{x1}), we arrive at
\begin{eqnarray}
&&\int_{0}^{1}dx\int_{0}^{1}2ydy\Big\{\Big([(k + q)^2 - m_{\chi^0}^2]x + (k^2 - m_{\tilde d_j}^2)(1 - x)\Big) y\nonumber\\&&\hspace{1.8cm}
 + (k^2 +p^2 +q^2 - m_{\tilde d_i}^2)(1 - y)\Big\}^{-3}\nonumber\\&&\hspace{1.8cm}
=\int_{0}^{1}dx\int_{0}^{1}2ydy\frac{1}{(k'^2 - J)^3}.
\end{eqnarray}

The needed expressions are

\begin{eqnarray}
&&(p + q)xy + q(1- y) = T\nonumber\\&&
k + T = k'\nonumber\\&&
J = -p^2xy + q^2[xy + (1-y)]^2 + 2p\cdot q[xy(xy - y)] + xy + (1 - y)\nonumber\\&&\hspace{1.0cm} +  m_{\chi^0}^2(1 - y) + m_{\tilde d_i}^2xy + m_{\tilde d_j}^2(1 - y)y.
\end{eqnarray}

Subsequently, we utilize D$-$dimensional spatial integration to perform relevant calculations on the denominators
\begin{eqnarray}
&&\int\frac{d^Dk^\prime}{(2\pi)^D}\frac{(k^2)^\alpha}{(k^2-R^2)^\beta}=i\frac{(-1)^{\alpha-\beta}}{(4\pi)^{\frac{D}{2}}}\frac{\Gamma(1+\frac{D}{2})\Gamma(\beta-\alpha-\frac{D}{2})}
{\Gamma(\frac{D}{2})\Gamma(\beta)(R^2)^{\beta-\alpha-\frac{D}{2}}}.
\end{eqnarray}

We use dimensional regularization to handle divergent terms, where $d = 4 - 2\epsilon$ and take the limit as d approaches 4. To obtain finite results, the divergent terms are canceled out by the modified minimal subtraction $(\overline{MS})$ scheme.

The final integral result we get is
\begin{eqnarray}
&&\mathcal{M}_{(a)} = \int_{0}^{1}dx\int_{0}^{1}2ydy \Big((\frac{- A_L B_L m_{\chi^0} - A_L B_R m_b x y - A_L B_R m_s x y + A_L B_R m_s y}{64\pi^2J})C_1 \gamma^5\nonumber\\&&
 + \frac{m_b A_L B_R C_1 x y}{64\pi^2J} + m_s C_1\frac{A_L B_R x y - A_L B_R y}{64\pi^2J} + \frac{m_{\chi^0} A_L B_L C_1}{64\pi^2J} + (L \rightarrow R)\Big).
\end{eqnarray}

We process all the diagrams in Fig.$\ref{N1}$ according to the aforementioned methods and further simplify them.

Due to the immense complexity of the calculations involved, we employ $Mathematica \ll HighEnergyPhysics`FeynCalc`$ package for analytical computations. We derive the Feynman amplitudes for all the diagrams. Subsequently, we perform summation operations and compute the squared modulus of the amplitudes, denoted as $|\mathcal{M}|^2$.

\section{Calculate the mass of SM-like Higgs boson}
The mass of the SM-like Higgs boson can be written as
\begin{eqnarray}
&&m_h=\sqrt{(m_{h}^0)^2+\Delta m_h^2},
\label{15}
\end{eqnarray}
where $m_{h}^0$ is the lightest CP-even Higgs boson mass at tree level. $\Delta m_h^2$ is the radiative correction for which an approximate expression including two-loop leading-log radiative corrections can be given as ~\cite{HiggsC1,HiggsC2,HiggsC3}
\begin{eqnarray}
&&\Delta m_h^2=\frac{3m_t^4}{4\pi^2 v^2}\Big[\Big(\tilde{t}+\frac{1}{2}\tilde{X}_t\Big)+\frac{1}{16\pi^2}\Big(\frac{3m_t^2}{2v^2}-32\pi\alpha_3\Big)\Big(\tilde{t}^2
+\tilde{X}_t \tilde{t}\Big)\Big],\nonumber\\
&&\tilde{t}=\log\frac{M_T^2}{m_t^2},\qquad\;\tilde{X}_t=\frac{2\tilde{A}_t^2}{M_T^2}\Big(1-\frac{\tilde{A}_t^2}{12M_T^2}\Big),
\label{eq16}
\end{eqnarray}
where $\alpha_3$ is the strong coupling constant, $M_T=\sqrt{m_{\tilde t_1}m_{\tilde t_2}}$ with $m_{\tilde t_{1,2}}$ denoting the stop masses, $\tilde{A}_t=A_t-(\mu+\frac{\lambda_Hv_S}{\sqrt{2}}) \cot\beta$ with $A_t=T_{u,33}$ being the trilinear Higgs stop coupling.


\begin{thebibliography}{33}
\vspace{3mm}
\bibitem{Higg1}ATLAS Publications, Phys. Lett. B {\bf 716} (2022) 1-29 [arXiv: 1207.7214].
\bibitem{NewH1}CMS Collaboration, Phys. Lett. B {\bf 716} (2012) 30-61 [arXiv: 1207.7235].
\bibitem{NewH2}J.~Ren, R.~Q.~Xiao, M.~Zhou, et al., JHEP {\bf 06} (2018) 090 [arXiv: 1706.05980].
\bibitem{NewH3}H.~Sun, Y.~J.~Zhou, JHEP {\bf 11} (2012) 127 [arXiv: 1211.6201].
\bibitem{NewH4}M.~Ibe, S.~Matsumoto, T.~T.~Yanagida, Phys. Rev. D {\bf 85} (2012) 095011 [arXiv: 1202.2253].
\bibitem{NewH5}J.~L.~Evans, M.~Ibe, S.~Shirai, et al., Phys. Rev. D {\bf 85} (2012) 095004 [arXiv: 1201.2611].
\bibitem{NewH6}T.~Moroi, K.~Nakayama, Phys. Lett. B {\bf 710} (2012) 159-163 [arXiv: 1112.3123].
\bibitem{FCNC}F.~Arco, S.~Heinemeyer, M.~J.~Herrero (2023) [arXiv: 2306.07958].
\bibitem{Higg2}R.~S.~Willey, H.~L.~Yu, Phys. Rev. D {\bf 26} (1982) 3086-3091.
\bibitem{Br1}L.~G.~Benitez-Guzm\'an, I.~Garc\'\i{}a-Jim\'enez, M.~A.~L\'opez-Osorio, et al., J. Phys. G {\bf 42} (2015) 085002 [arXiv: 1506.02718].
\bibitem{Br2}J.~I.~Aranda, G.~Gonz\'alez-Estrada, J.~Monta\~no, et al., J. Phys. G {\bf 47} (2020) 125001 [arXiv: 2009.07166].
\bibitem{Br3}G.~Blankenburg, J.~Ellis, G.~Isidori, Phys. Lett. B {\bf 712} (2012) 386-390 [arXiv: 1202.5704].
\bibitem{Br4}D.~Barducci, A.~J.~Helmboldt, et al., JHEP {\bf 12} (2017) 105 [arXiv: 1710.06657].
\bibitem{mu1}C.~X.~Liu, H.~B.~Zhang, J.~L.~Yang, et al., JHEP {\bf 04} (2020) 002 [arXiv: 2002.04370].
\bibitem{mu2}H.~B.~Zhang, T.~F.~Feng, G.~H.~Luo, et al., JHEP {\bf 07} (2013) 069 [arXiv: 1305.4352].
\bibitem{zwz1}S.~M.~Zhao, T.~F.~Feng, X.~X.~Dong, et al., Nucl. Phys. B {\bf 910} (2016) 225-239 [arXiv: 1603.09505].
\bibitem{zwz2}S.~Davidson, S.~F.~King, Phys. Lett. B {\bf 445} (1998) 191-198 [arXiv: 9808296].
\bibitem{U1x2}S.~M.~Zhao, L.~H.~Su, X.~X.~Dong, et al., JHEP {\bf 03} (2022) 101.
\bibitem{U1x1}B.~Yan, S.~M.~Zhao, T.~F.~Feng, et al., Nucl. Phys. B \textbf{975} (2022) 115671 [arXiv: 2011.08533].
\bibitem{U1x3}S.~M.~Zhao, T.~F.~Feng, M.~J.~Zhang, et al., JHEP {\bf 02} (2020) 130 [arXiv: 1905.11007].
\bibitem{U1x4}M.~Y.~Liu, S.~M.~Zhao, S.~Gao, et al., (2024) [arXiv: 2405.00961].
\bibitem{Right1}Y.~T.~Wang, S.~M.~Zhao, T.~T.~Wang., et al., Phys. Rev. D {\bf 106} (2022) 055044 [arXiv: 2207.01770].
\bibitem{Right2}S.~M.~Zhao, G.~Z.~Ning, J.~J.~Feng, et al., Nucl. Phys. B {\bf 969} (2021) 115469.
\bibitem{Higg3}R.~El-Kosseifi, J.~L.~Kneur, G.~Moultaka, et al., Eur. Phys. J. C {\bf 82} (2022) 657 [arXiv: 2202.06919].
\bibitem{TREE}L.~H.~Su, S.~M.~Zhao, X.~X.~Dong, et al., Eur. Phys. J. C {\bf 81} (2021) 433 [arXiv: 2012.04824].
\bibitem{QFV}F.~Abu-Ajamieh, M.~Frasca, S.~K.~Vempati, JHEP \textbf{02} (2016) 075010 [arXiv:2305.17362].
\bibitem{LFV11}R.~K.~Barman, P.~S.~B.~Dev, A.~Thapa, Phys. Rev. D \textbf{107} (2023) 075018 [arXiv:2210.16287].
\bibitem{LFV22}R. Aaij, et al., Eur. Phys. J. C \textbf{78} (2028) 1008 [arXiv:1808.07135].
\bibitem{LFV33}ATLAS Publications, JHEP \textbf{07} (2023) 166 [arXiv:2302.05225].
\bibitem{LFV44}M.~Badziak, G.~Grilli di Cortona, M.~Tabet, et al., JHEP \textbf{10} (2021) 181 [arXiv:2107.09708].
\bibitem{LFV1}J.~J.~Zhang, M.~He, X.~G.~He, et al., JHEP \textbf{02} (2019) 007 [arXiv:1807.00921].
\bibitem{LFV2}CMS Collaboration, Phys. Rev. D \textbf{108} (2023) 072004 [arXiv:2305.18106].
\bibitem{LFV3}ATLAS Collaboration, JHEP \textbf{07} (2023) 166 [arXiv:1907.05900].
\bibitem{BB1}J.~H.~Garcia, M.~Nebot, F.~Rajec, et al., JHEP \textbf{02} (2020) 147 [arXiv: 1907.05900].
\bibitem{BB2}G.~Blankenburg, J.~Ellis, G.~Isidori, Phys. Lett. B  \textbf{712} (2012) 386-390 [arXiv: 1202.5704].
\bibitem{BB3}D.~Barducci, A.~J.~Helmboldt, JHEP \textbf{12} (2017) 105 [arXiv:1710.06657].
\bibitem{NMSSM1}J.J.~Cao, D.W.~Li, L.L.~Shang, P.W.~Wu, et al., JHEP {\bf12} (2014) 026 [arXiv:1409.8431].
\bibitem{NMSSM2}J.J.~Cao, X.F.~Guo, L.L.~Shang, Y.L.~He, et al., Phys. Rev. D {\bf95} (2017) 116001 [arXiv:1612.08522].
\bibitem{Du1}G.~B\'elanger, J.~Da Silva, H.~M.~Tran, Phys. Rev. D {\bf 95} (2017) 115017 [arXiv: 1703.03275].
\bibitem{Du2}V.~Barger, P.~Fileviez~Perez, S.~Spinner, Phys. Rev. Lett. {\bf 102} (2009) 181802 [arXiv: 0812.3661].
\bibitem{Du3}P.~H.~Chankowski, S.~Pokorski, J.~Wagner, Eur. Phys. J. C {\bf 47} (2006) 187
\bibitem{Higgs coup}S.~M.~Zhao, X.~Wang, X.~X.~Dong, et al., Symmetry {\bf 14} (2022) 2153 [arXiv: 2209.07094].
\bibitem{HiggsM}D.~de Florian, et al., doi:10.23731/CYRM-2017-002 [arXiv:1610.07922].
\bibitem{pdg}R.L. Workman, et al,. (Particle Data Group), Prog. Theor. Exp. Phys. \textbf{2022} (2022) 083C01.
\bibitem{g-21}\'A.~S.~de Jesus, F.~S.~Queiroz, J.~W.~F.~Valle, et al., (2024) [arXiv: 2312.03851].
\bibitem{g-22}D.~P.~Aguillard et al., Phys. Rev. Lett. {\bf 131} (2023) 161802 [arXiv: 2308.06230].
\bibitem{limit1}P.~Cox, C~.C.~Han, T.~T.~Yanagida, Phys. Rev. D \textbf{104} (2021) 075035 [arXiv:2104.03290].
\bibitem{limit2}M.~V.~Beekveld, W.~Beenakker, M.~Schutten, et al., SciPost Phys. \textbf{11} (2021) 3, 049 [arXiv:2104.03245].
\bibitem{limit3}M.~Chakraborti, L.~Roszkowski, S.~Trojanowski, JHEP \textbf{05} (2021) 252 [arXiv:2104.04458].
\bibitem{limit4}F.~Wang, L.~Wu, Y.~Xiao, et al., Nucl. Phys. B \textbf{970} (2021) 115486 [arXiv:2104.03262].
\bibitem{limit5}M.~Chakraborti, S.~Heinemeyer, I.~Saha, Eur. Phys. J. C \textbf{81} (2021) 12, 1114 [arXiv:2104.03287].
\bibitem{limit6}M.~Endo, K.~Hamaguchi, S.~Iwamoto, et al., JHEP \textbf{07} (2021) 075 [arXiv:2104.03217].
\bibitem{MDM1} D.~P.~Aguillard, et al. Phys. Rev. Lett. {\bf131} (2023) 161802.
\bibitem{MDM2} A.~Datta, D.~Marfatia, L.~Mukherjee, Phys. Rev. D. {\bf109} (2024) L031701.
\bibitem{Finally1}G.~Blankenburg, J.~Ellis, G.~Isidori, Phys. Lett. B {\bf 712} (2012) 386.
\bibitem{Finally2}R.~Harnik, J.~Kopp, J.~Zupan, JHEP {\bf 03} (2013) 026.
\bibitem{Du4}B.~L.~Yang, Quantum Field Theory: From Operators to Path Integrals, World Scientific Publishing Co. (1988).
\bibitem{HiggsC1}M. Carena, J. R. Espinosaos, C. E. M. Wagner, and M. Quir, Phys. Lett. B \textbf{355} (1995) 209.
\bibitem{HiggsC2}M. Carena, M. Quiros, and C. E. M. Wagner, Nucl. Phys. B \textbf{461} (1996) 407.
\bibitem{HiggsC3}M. Carena, S. Gori, N.R. Shah, and C. E. M. Wagner, J. High Energy Phys. {\bf03} (2012) 014.
\end{thebibliography}
\end{document}